\documentclass[prl,twocolumn,floatfix,superscriptaddress,amsmath,citeautoscript,aps,longbibliography]{revtex4-2}

\usepackage{mathtools}
\usepackage{dcolumn,bm,graphicx,xcolor,booktabs,microtype,afterpage} 
    \graphicspath{{Figures/}}
\usepackage[charter,greekuppercase=italicized]{mathdesign}
\usepackage[toc]{appendix}
\usepackage{physics}
\usepackage[normalem]{ulem}

\renewcommand{\figurename}{Fig.}
\renewcommand{\tablename}{Table}

\makeatletter\renewcommand{\fnum@figure}[1]{\figurename~\thefigure.}\makeatother
\makeatletter\renewcommand{\fnum@table}[1]{\tablename~\thetable.}\makeatother
\newcount\hh \newcount\mm
\hh=\time \divide\hh by 60
\mm=\hh \multiply\mm by 60 \mm=-\mm
\advance\mm by \time
\def\now{\number\hh:\ifnum\mm<10{}0\fi\number\mm}
\usepackage[colorlinks,plainpages=false,linkcolor=blue,urlcolor=blue,citecolor=blue,pdfpagemode=UseNone,pdfstartview=FitBH]{hyperref}

\newcommand{\mh}{\mathcal{H}}

\newcommand{\be}{\begin{eqnarray}}
\newcommand{\ee}{\end{eqnarray}}

\newcommand{\Df}{D_\mathrm{f}}

\DeclareMathOperator{\IPR}{IPR}

\newcommand{\pcsadd}{Center for Theoretical Physics of Complex Systems, Institute for Basic Science, Daejeon 34126, Republic of Korea}
\newcommand{\ustadd}{Basic Science Program, Korea University of Science and Technology, Daejeon 34113, Republic of Korea}

\newcommand{\sect}[1]{\textit{#1}---}

\begin{document}

\title{Localization--non-ergodic transition in controllable-dimension fractal networks from diffusion-limited aggregation}

\author{Oleg I. Utesov}
    \thanks{utiosov@gmail.com}
    \affiliation{\pcsadd}
    \affiliation{Department of Physics, Korea Advanced Institute of Science and Technology, Daejeon 34141, Republic of Korea}

\author{Alexei Andreanov}
    \affiliation{\pcsadd}
    \affiliation{Center for Trapped Ions Quantum Science, Institute for Basic Science, Daejeon 34126, Republic of Korea}
    \affiliation{\ustadd}

\author{Tomasz Bednarek}
    \affiliation{\pcsadd}
    \affiliation{Institute of Physical Chemistry, Polish Academy of Sciences, Warsaw 01-224, Poland} 

\author{Alexandra Siklitskaya}
    \affiliation{Institute of Physical Chemistry, Polish Academy of Sciences, Warsaw 01-224, Poland} 

\author{Sergei V. Koniakhin}
    \thanks{kon@ibs.re.kr}
    \affiliation{\pcsadd}
    \affiliation{\ustadd}
    \affiliation{Russian Quantum Center, Skolkovo, Moscow 121205, Russian Federation}

\begin{abstract}
    Our study connects the physics of disordered integer-dimensional systems and regular self-similar objects by studying spectral properties of fractal agglomerates with tunable dimension.
    The latter is controlled by parameter \(\alpha\) of the algorithm that generates the agglomerates.
    We consider the nearest-neighbor tight-binding model on the agglomerates embedded in 2D and 3D, and observe that all eigenstates are localized in the 2D case, whereas in the 3D case, there is a localization--non-ergodic transition upon increasing \(\alpha\),
    i.e., going from sparse to dense fractals: a sub-extensive number of critical states emerge in the spectrum at a certain critical value of \(\alpha\).
    The complex geometry of the agglomerates is also responsible for a peculiar hierarchy of compact localized states and singularities in the density of states, which are typical for ordered fractals.    
\end{abstract}

\maketitle

\sect{Introduction}
The phenomenon of localization of elementary excitations in disordered quantum systems is well-known for a long time~\cite{anderson1958absence,evers2008anderson,abrahams201050}.
In a tight-binding model with on-site potential disorder, there is a metal-insulator transition in three spatial dimensions~\cite{abrahams1979,slevin2014critical}, whereas for the 2D case, all eigenstates are localized even for infinitesimal disorder strength (a phenomenon referred to as weak localization)~\cite{gorkov1979}.
Importantly, the Anderson model~\cite{anderson1958absence} can be generalized and formulated for various random matrix ensembles, which allows studying generic spectral properties of localization~\cite{mehta2004random}.
Moreover, the emergent field of many-body localization studies the interplay between disorder and interactions, and the corresponding problems can be treated in terms of hopping in a complicated Hilbert space of many-particle states~\cite{gornyi2005,basko2006,aleiner2010finite,tikhonov2021from,roy2019percolation,sierant2023universality}.
This highlights the importance of understanding single-particle localization on complex networks.

The so-called geometrical disorder represents another type of disorder that can drastically affect the spectral properties of the model.
One of the most studied models of this type is quantum percolation~\cite{de1959amas,kirkpatrick1972localized,pant1980quantum,shapir1982,schubert2005localization,bhattacharjee2025anderson}, where a random fraction of sites is removed from the tight-binding network.
The important results from Ref.~\onlinecite{shapir1982} are: 
(i) for 3D systems, there is a metal-insulator transition at a critical site dilution which is above the classical percolation threshold, and
(ii) localized and delocalized modes can coexist in the spectrum without a mobility edge due to the complex geometry of the diluted network.
Below, we refer to this phenomenon as the \emph{spectrum intermixture}.
It was understood recently that interference conditions can be realized at the surfaces of complex geometric structures~\cite{roentgen2018compact}, resulting in the emergence of the so-called compact localized states (CLS) typical for flat-band systems~\cite{derzhko2015strongly,leykam2018artificial,rhim2021singular,danieli2024flat}.
The complexity of the tight-binding network can also be encoded by its fractal structure.
For regular fractals, such as the Serpinsky gasket, exact self-similarity leads to peculiar spectral properties with coexisting localized and delocalized states, as well as non-ergodic multifractal states ~\cite{rammal1983nature,domany1983solutions,vanVeen2016, salvati2026emergence}.
In particular, in Ref.~\onlinecite{domany1983solutions} it was shown that the fraction of extended states vanishes exponentially with the structure iteration order.
The Anderson model on regular fractals was also studied~\cite{sticlet2016attractive,shou2024spectrum}.

The focus of this letter is objects that combine both disorder and self-similarity properties.
We consider nearest-neighbor tight-binding models on ensembles of random fractal agglomerates.
Notably, we proposed an algorithm for generating agglomerates that allows to tune their fractal dimensions over a relatively broad range.
For the agglomerates with dimensions \(\Df \in (1.3, 1.9)\) embedded in 2D, we observed a trivial result that all eigenstates are localized.
In contrast, in the 3D case, we observed a localization--non-ergodic transition upon variation of the fractal dimension, \(\Df \in (1.6, 2.8) \).
In the non-ergodic phase, a sub-extensive number of critical (fractal) modes are immersed in the otherwise localized spectrum.
These, usually elusive states, naturally emerge in our nearest-neighbor model and span a broad range of energies.
Moreover, for all clusters, we observed geometry-induced singularities in the density of states (DOS), similar to those discussed in Refs.~\cite{kirkpatrick1972localized,domany1983solutions,schubert2005localization,roentgen2018compact}.
The corresponding compact localized states mostly reside on the dendrites of the agglomerates and have a nontrivial hierarchy.
Our study bridges the physics of disordered integer-dimension crystals and regular self-similar fractal structures and can be important for understanding properties of real-life irregular fractals.

\sect{Structures and Dimensions}
Our algorithm for the agglomerates' generation is inspired by the so-called cluster-cluster (C-C) and particle-cluster (P-C) aggregation mechanisms~\cite{brasil2000evaluation,brasil2001numerical} in the discretized space (see Supplemental Material for details and Ref.~\onlinecite{tomchuk2023models} for review).
For the former, on each step, two agglomerates of the same size collide and stick together, which doubles the size of the agglomerate. 
For the latter, one particle randomly sticks to the existing structure.
As a result, P-C agglomerates are denser in comparison with relatively sparse C-C ones.
This observation can be quantified using the fractal dimension~\cite{brasil2000evaluation}, defined as
\begin{gather} 
    \label{eq:dfr}
    \Df = \lim_{N \to \infty} \frac{\ln{N}}{\ln{\sqrt{\langle R^2_i \rangle}}},
\end{gather}
i.e., describing how the number of sites \(i\) scales with the typical linear size of the agglomerate \(\sqrt{\langle R^2_i \rangle}\), where we calculate the average distance between all the sites and the agglomerate center of mass. 

\begin{figure}
    \centering
    \includegraphics[width=0.49\columnwidth]{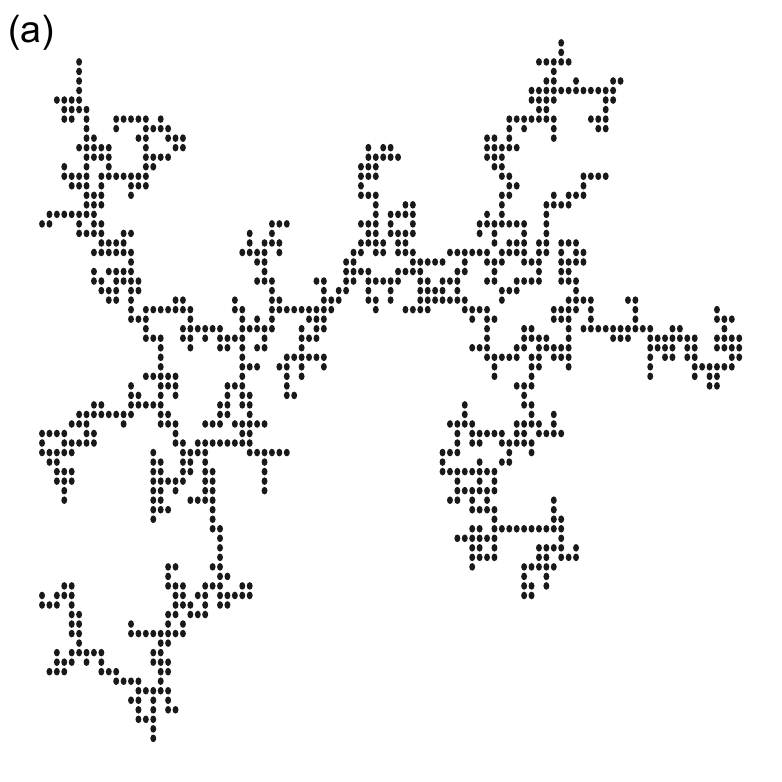}
    \includegraphics[width=0.49\columnwidth]{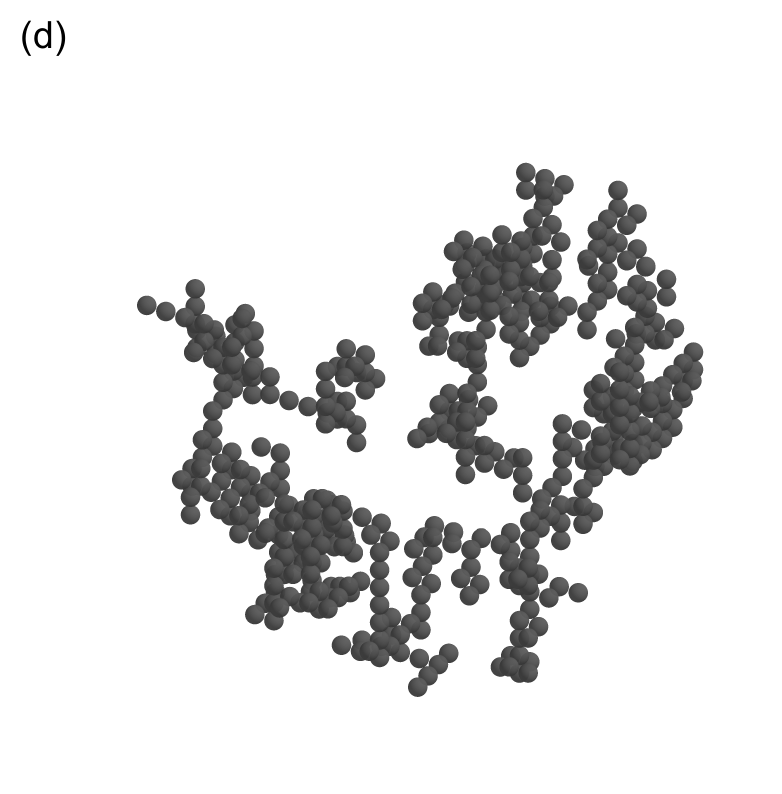}
    \includegraphics[width=0.49\columnwidth]{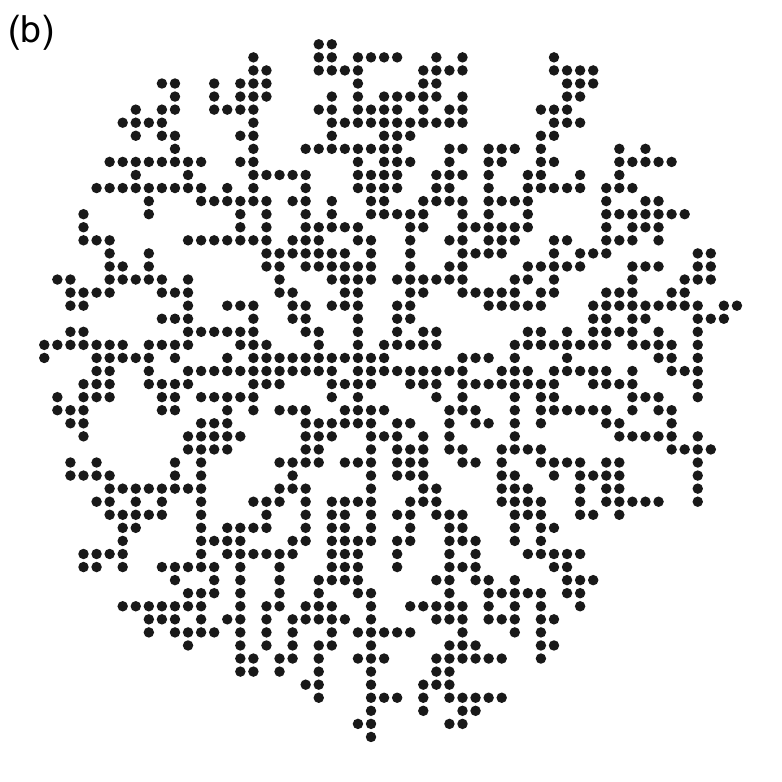}
    \includegraphics[width=0.49\columnwidth]{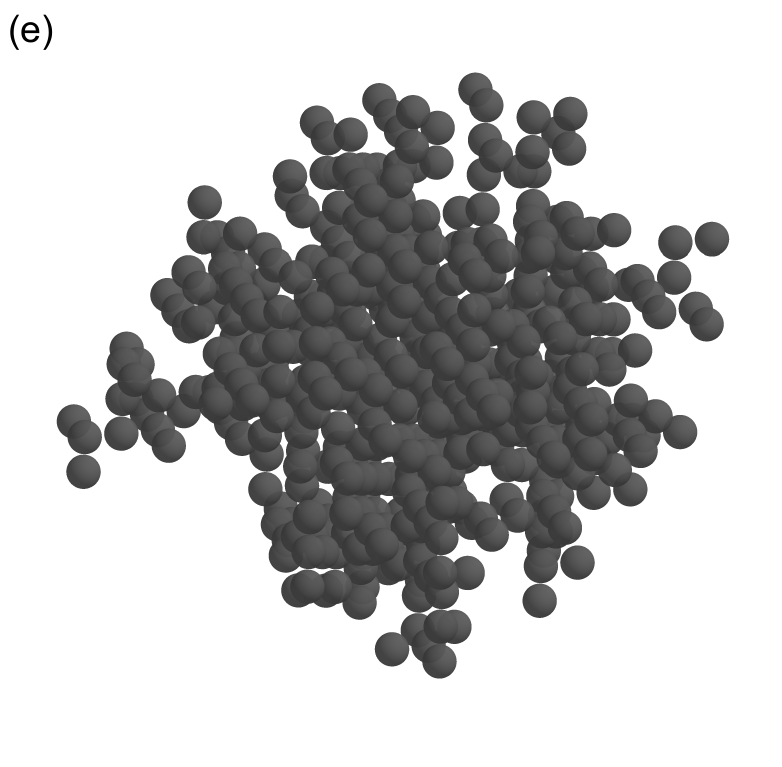}
    \includegraphics[width=0.49\columnwidth]{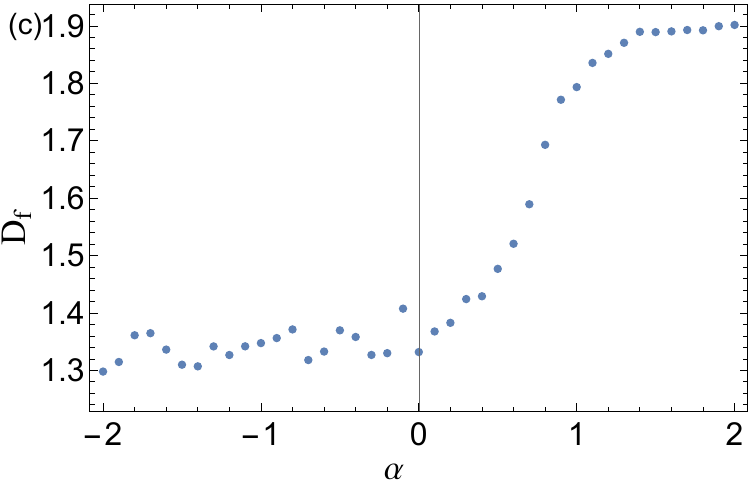}
    \includegraphics[width=0.49\columnwidth]{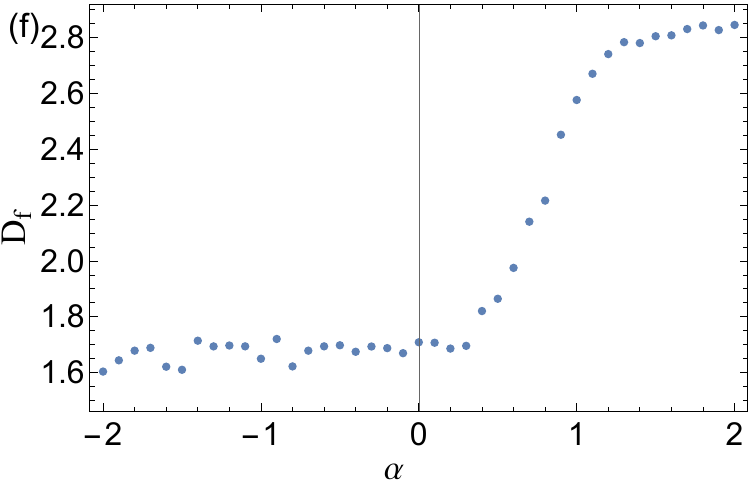}
    \caption{
        Fractal agglomerates in 2D (left) and 3D (right).
        (a) $N=1024$ cluster-cluster agglomerate in 2D space.
        (b) $N=1024$ particle-cluster agglomerate in 2D.
        (c) Parameter $\alpha$ defines fractal dimensions of the 2D agglomerates, allowing for their tuning in a certain range, interpolating between C-C and P-C models.
        (d) $N=512$ cluster-cluster agglomerate in 3D.
        (e) $N=512$  particle-cluster agglomerate in 3D.
        (f) The same as (c) but for 3D agglomerates.
        In (c) and (f), we used averaging over 256 samples for each size $N = 2^{7 \text{---} 12}$ and scaling relation~\eqref{eq:dfr}. 
        Note that $\alpha \in (-2,2)$ practically covers the entire range of available fractal dimensions in both cases.
    }
    \label{fig:fractals}
\end{figure}

Apparently, P-C agglomerates have higher \(\Df\) than the C-C ones. 
A natural question: ``can we interpolate between these two models in a more-or-less controllable way?'' emerges. 
To do so, we introduce the aggregation kernel~\cite{van1988scaling,schmitt2000multiple} \(K(M_1,M_2) = M_1^\alpha M_2^\alpha\) (which is the reaction rate in the Smoluchowski equation) with the crucial parameter \(\alpha\) governing the probability of two agglomerates of sizes \(M_{1,2}\) to coagulate.
To create an agglomerate with \(N\) particles, we start with \(N\) monomers, then, on each step, choose two of them according to the kernel above, and collide in the discretized space with random relative orientation.
Further details are discussed in Supplemental Material.

Notably, for $\alpha \to -\infty$, in each step we choose the smallest aggregates, so that their size is doubled, hence the C-C model is reproduced, whereas for $\alpha \to +\infty$, on each step the largest agglomerate and a monomer are chosen and we arrive at the P-C one. 
Several examples of resulting agglomerates as well as fractal dimensions as functions of \(\alpha\) are presented in Fig.~\ref{fig:fractals}.
For the 2D agglomerates, we have roughly \(\Df \in (1.3, 1.9)\), and in 3D: \(\Df \in (1.6, 2.8)\).
This means that in 3D, we can tune from very sparse dendrite-like to dense globular-like structures, which can have very different properties.

\sect{Tight-binding model and spectral properties}
We consider a clean nearest-neighbor tight-binding model on an agglomerate, i.e., 
\begin{gather}
    \label{eq:tbm}
    \mh = \sum_{\langle i,j\rangle} (\ketbra{i}{j} + \ketbra{j}{i}),
\end{gather}
where all energies are measured in units of the hopping parameter.
Therefore, the Hamiltonian is equivalent to the adjacency matrix of the agglomerate.
In the discrete space, the bonds connect sites with the coordinates difference \(\pm (1,0)\) and \((\pm 1,0,0)\) (plus permutations of Cartesian directions) for the 2D and 3D cases, respectively.

\begin{figure*}
    \centering
    \includegraphics[width=0.28\linewidth]{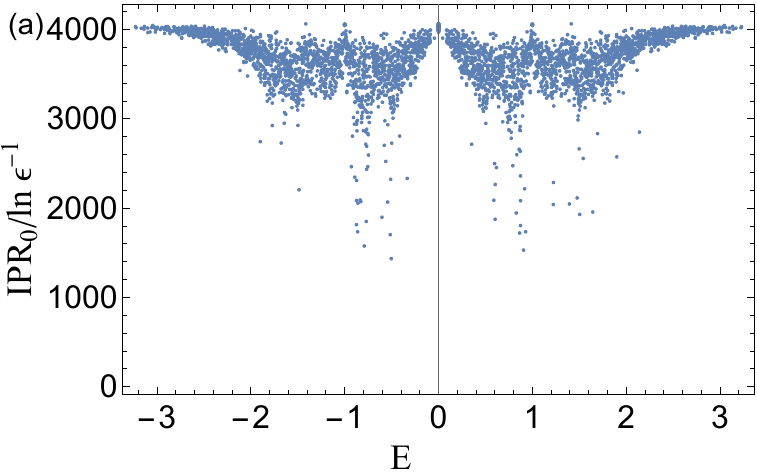}
    \hspace{1cm}
    \includegraphics[width=0.26\linewidth]{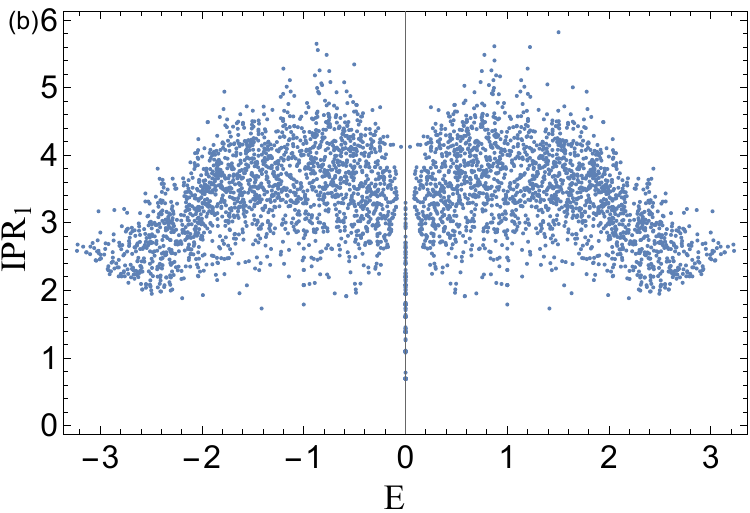}
    \hspace{1cm}
    \includegraphics[width=0.27\linewidth]{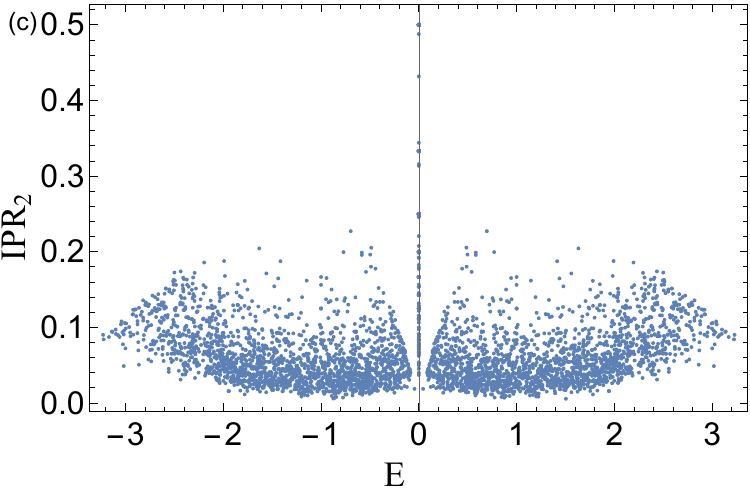}
    \includegraphics[width=0.28\linewidth]{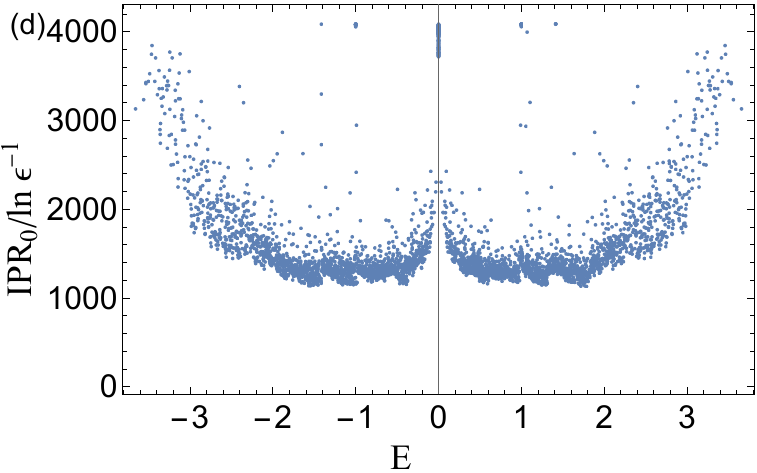}
    \hspace{1cm}
    \includegraphics[width=0.26\linewidth]{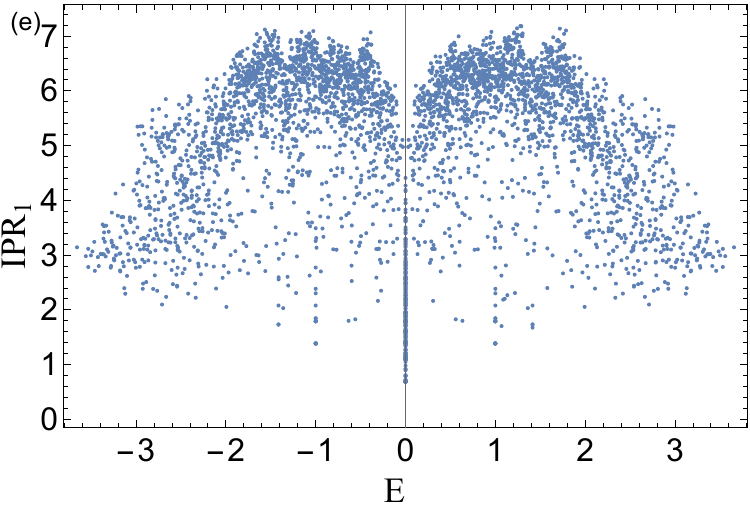}
    \hspace{1cm}
    \includegraphics[width=0.27\linewidth]{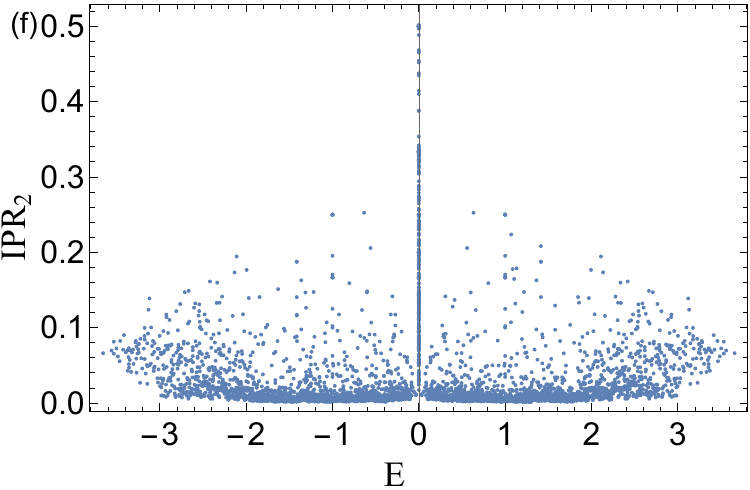}
    \caption{
       IPRs of the eigenstates of the tight-binding model on cluster-cluster and particle-cluster fractal agglomerates.
        Top row: 2D C-C model with \(N=2^{12}\) sites.
        Most of the states are localised with \(\IPR_0 \approx N \ln{(1/\epsilon)}\) .
        Only a few modes occupy \(\sim N/2\) sites as visible from \(\IPR_1\) and \(\IPR_2\).
        Bottom row: 3D P-C model with \(N=2^{12}\) sites.
        Based on \(\IPR_0\) most states occupy \(>N/2\) sites, their \(\IPR_1 \in (6,7)\) (for a homogeneously spread wave-function one would have $\approx 8$) and \(\IPR_2\) is almost zero $\sim 1/N$.
        The spread of \(\IPR_1\) values suggests a strong mix of localized and non-localized modes for most eigenenergies.
        }
    \label{fig:exact}
\end{figure*}

We characterize an eigenstate of the model~\eqref{eq:tbm} by its energy \(E\) and Inverse Participation Ratios (IPRs)~\cite{evers2008anderson,he2022persistent}:
\begin{align}
    \IPR_0(\nu) &= - \sum_i \ln{(|\psi^\nu_i|^2 + \epsilon)}, \\
    \IPR_1(\nu) &= - \sum_i |\psi^\nu_i|^2 \ln{(|\psi^\nu_i|^2 + \epsilon)}, \\
    \IPR_2(\nu) &= \sum_i |\psi^\nu_i|^4,
\end{align}
where $\nu$ is the index of the eigenstate; we set $\epsilon = 10^{-16} \ll 1/N$ for regularization of $\IPR_{0,1}$. 
The probes above can be motivated as follows.
For a given state, $\IPR_0$ roughly indicates the number of sites where its amplitude is negligible.
For instance, for the state localized on a single site, $\IPR_0 = (N-1) \ln{10^{16}}$.
If the state is localized on \(M\) sites $\IPR_0 \approx (N-M) \ln{10^{16}}$.
The corresponding values of $\IPR_1$ and $\IPR_2$ are $\sim \ln{M}$ and $\sim 1/M$, respectively~\cite{he2022persistent}. 
However, the true nature of states can only be revealed by studying IPRs' scaling with the system size \(N\), which defines their fractal exponents $\tau$~\cite{evers2008anderson}.
For extended states $\tau=1$, for fractal (critical) $0<\tau<1$, and for localized ones $\tau=0$.
One can imagine that wave-functions near a particular energy spread over $N^\tau$ sites homogeneously, then apparently $N-\IPR_0/\ln{\epsilon} \sim N^\tau$, $\IPR_1 \sim \tau \ln{N}$, and $\IPR_2 \sim 1/N^\tau$.
The exponent \(\tau\) could also depend on the IPR used, indicating multifractal eigenstates~\cite {evers2008anderson}.


We first analyze the spectral properties for moderate sizes, where exact diagonalization of the Hamiltonian~\eqref{eq:tbm} is feasible.
Our findings are shown in Fig.~\ref{fig:exact} and have several important features:
(i) the spectra typically span the \((-4,4)\) range,
(ii) the spectra are mixed: 
for almost all energies, one observes a wide spread of participation numbers, and
(iii) there is a macroscopic degeneracy at \(E=0\) 

\begin{figure*}
    \centering
    \includegraphics[width=0.3\linewidth]{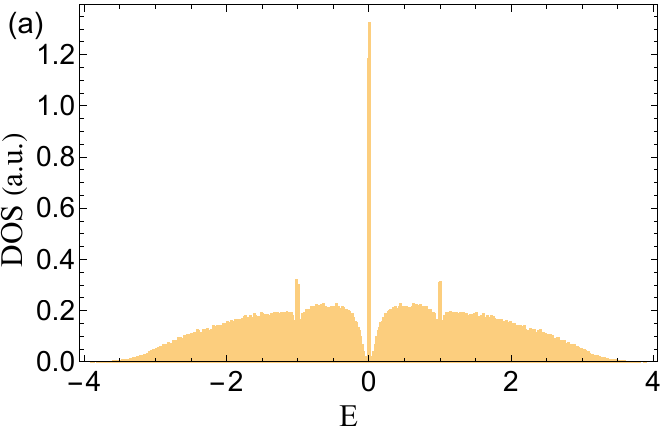}
    \hspace{0.5cm}
    \includegraphics[width=0.3\linewidth]{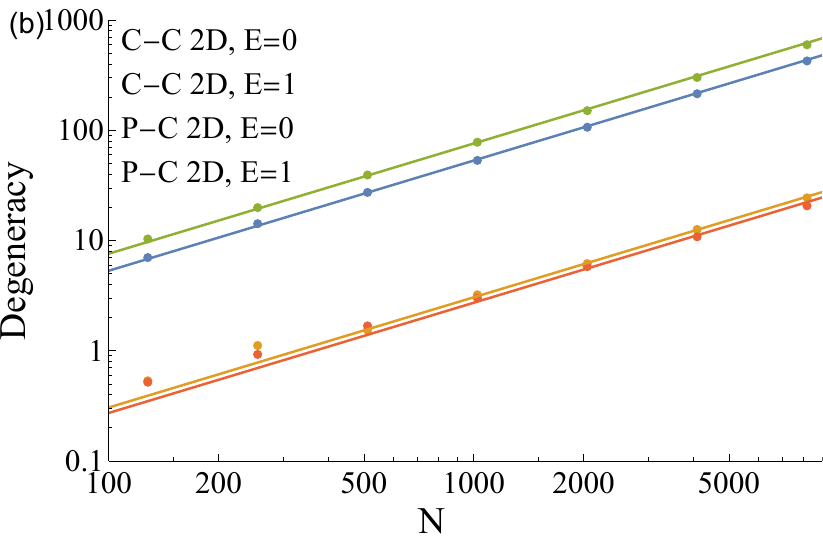}
    \hspace{0.5cm}
    \includegraphics[width=0.3\linewidth]{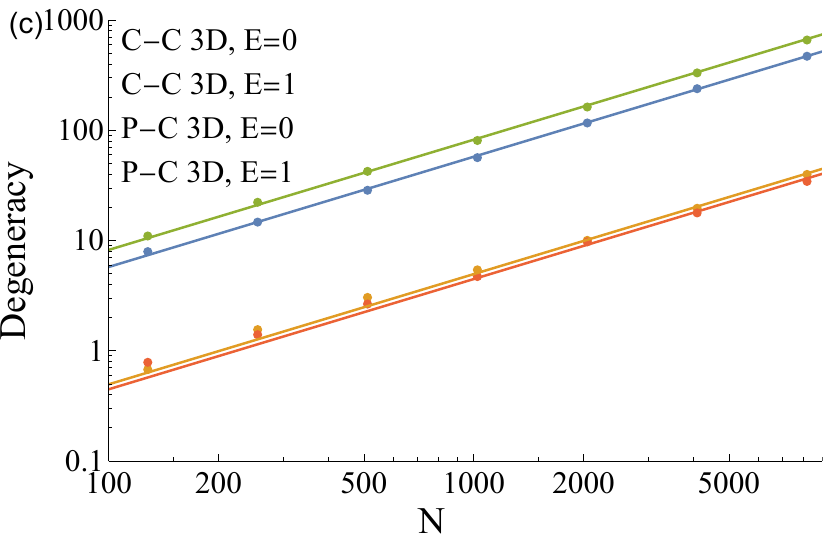}
    \includegraphics[width=0.23\linewidth]{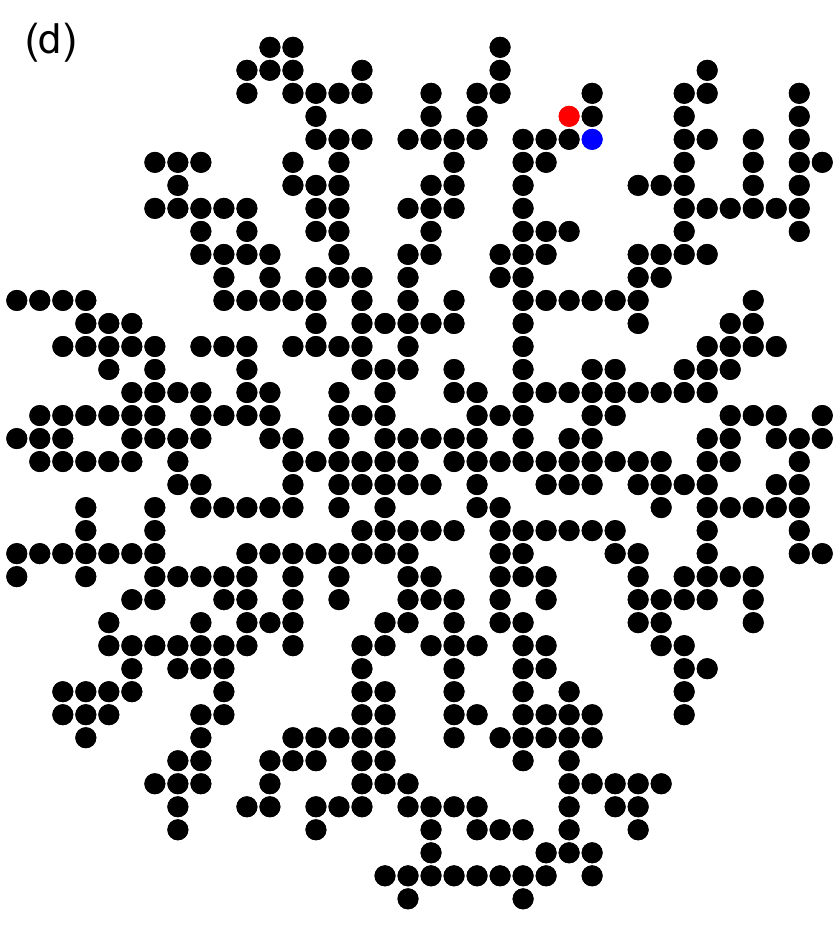}
    \hspace{1.5cm}
    \includegraphics[width=0.23\linewidth]{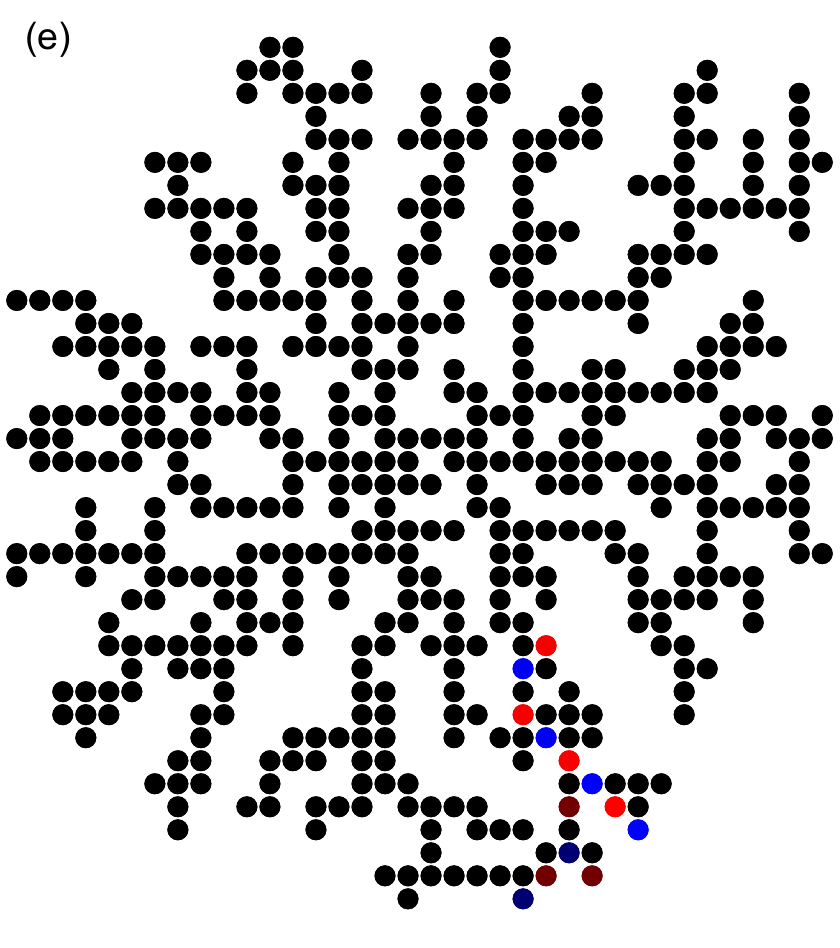}
    \hspace{1.5cm}
    \includegraphics[width=0.23\linewidth]{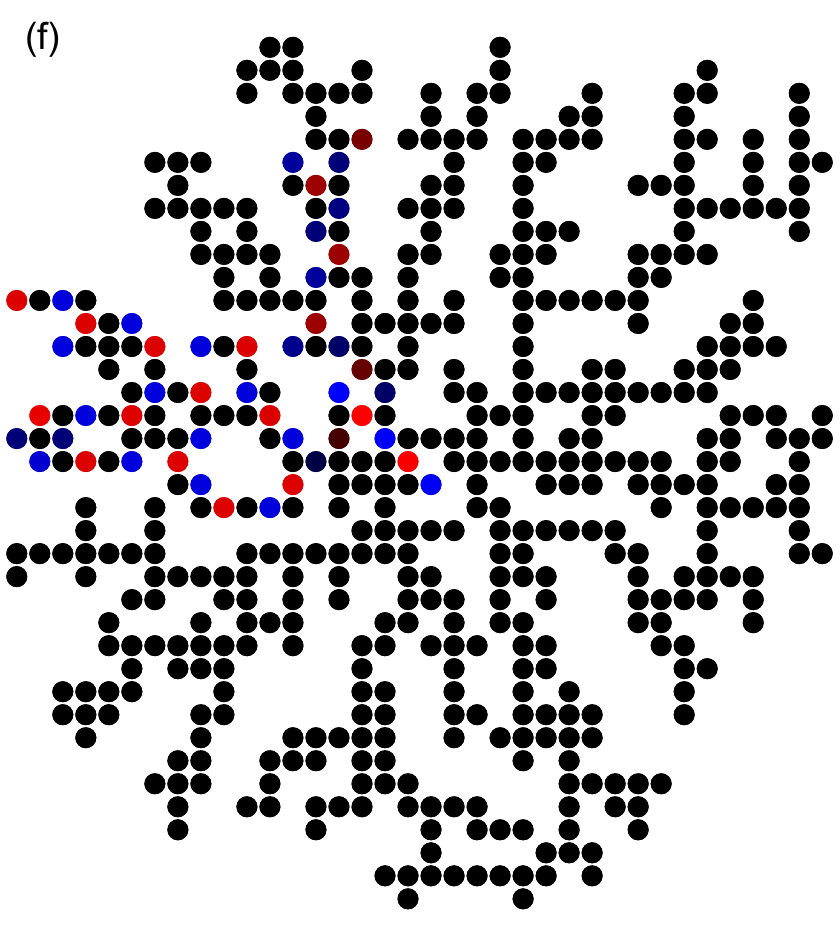}
    \caption{
         (a) Density of states for the tight-binding model on fractal 3D agglomerates with \(\alpha=0\) and $N=2^{12}$ averaged over \(64\) samples.
         (b) Scaling of the number of \(E=0\) and \(E=1\) eigenstates with \(N\) for 2D C-C and P-C agglomerates.
         (c) Scaling of the number of \(E=0\) and \(E=1\) eigenstates with \(N\) for 3D C-C and P-C agglomerates.
         (d)-(f) Examples of the \(E=0\) eigenstates for 2D P-C agglomerate.
         Red and blue colors mark the sign of the amplitudes, while the color intensity corresponds to their absolute values.
         Black color indicates zero amplitude of the wave function. 
    }
    \label{fig:degen}
\end{figure*}

We continue with the discussion of DOS, which is remarkably similar for 2D and 3D fractals.
The DOS computed for 3D and \(\alpha=0\) is shown in Fig.~\ref{fig:degen}(a) (see Supplementary Material for the 2D case).
On top of the relatively smooth part, one can see the singular peaks at \(E=0, \pm 1\), whose origin is similar to those appearing in quantum percolation discussed in Ref.~\onlinecite{kirkpatrick1972localized} and in regular self-similar fractals~\cite{rammal1983nature,domany1983solutions}.
They are inherent to the considered agglomerates and are related to the underlying geometrical structure and destructive interference~\cite{roentgen2018compact}.
We examine these degeneracies for 2D and 3D, C-C and P-C agglomerates of sizes $N = 2^{7 \text{---} 13}$.
Remarkably, these degeneracies are extensive, i.e., $\propto N$, see Figs.~\ref{fig:degen}(b) and (c).
Examples of the \(E=0\) eigenstates of the 2D P-C fractal are shown in Figs.~\ref{fig:degen}(d)-(f).
They are made of \emph{real} alternating in space on-site amplitudes separated by trivial $\psi_i=0$ sites.
The simplest such eigenstates reside on just two sites, while more complex patterns spanning multiple sites are also present.
Such spatially compact eigenstates are similar to CLS  in flat-band systems~\cite{derzhko2015strongly,leykam2018artificial,rhim2021singular,danieli2024flat}.
Thus, there is a complex hierarchy of degenerate CLS-like states.
Eigenstates for $E=\pm 1$ are presented in the End Matter, see Fig.~\ref{fig:wfe1}.
We conclude that our fractal agglomerates' complex geometry leads to a peculiar density of states similar to regular self-similar objects.

\sect{2D agglomerates and localisation}
We use the IPR to analyze spectral properties of the eigenstates of~\eqref{eq:tbm} in two spatial dimensions.
On general grounds, one can expect localisation of all states in this case, since it is a model with geometrical disorder embedded in 2D with \(\Df < 2\) for all structures.
To confirm this intuition, we performed exact diagonalization for sizes \(N \leq 2^{13}\), selecting the least localized eigenstates, with maximal \(\IPR_1\) and minimal \(\IPR_2\), for each realization.

\begin{figure}
    \centering
    \includegraphics[width=0.47\linewidth]{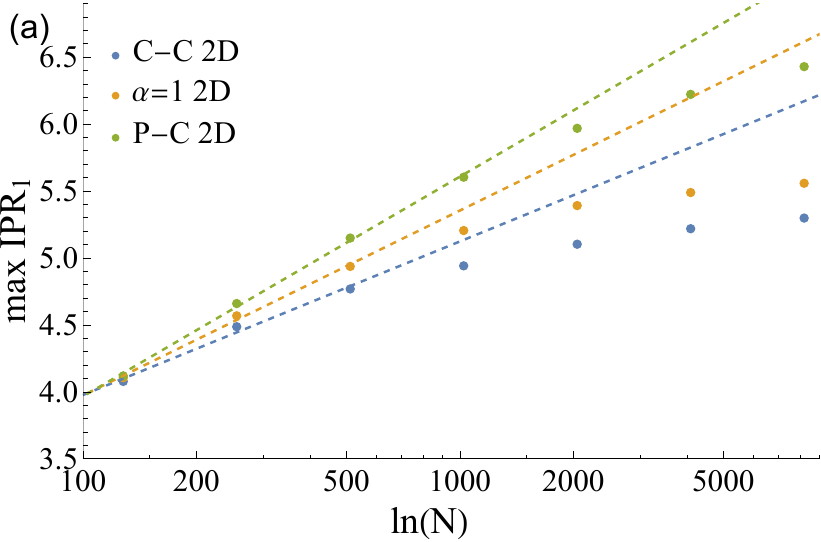}
    \includegraphics[width=0.47\linewidth]
    {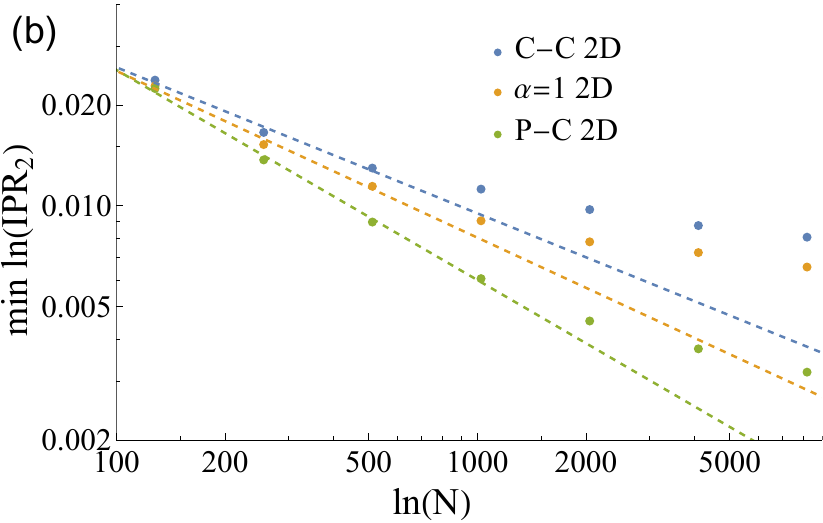}
    \caption{
        IPR of most delocalized states for 2D agglomerates with C-C (\(\alpha \to -\infty\)), \(\alpha=1\), and P-C (\(\alpha \to +\infty\)).
        (a) Maximum \(\IPR_1\) averaged over \(64\) realizations per size.
        (b) Minimum \(\IPR_2\) averaged over \(64\) realizations per size.
        In both panels, dashed lines are guides for the eyes representing IPRs of delocalized states.
        Clear saturation of IPRs is observed, indicating localization of all states in 2D.
        }
    \label{fig:ipr2D}
\end{figure}

The exact diagonalization results for C-C ($\alpha \to -\infty$), intermediate \(\alpha=1\) case, and P-C ($\alpha \to +\infty$) agglomerates are shown in Fig.~\ref{fig:ipr2D}.
Both \(\IPR_1\) and \(\IPR_2\) display an apparent tendency to saturation with the agglomerates' size, indicating localization of the corresponding ``most delocalized'' states and confirming our intuition presented above.

Our results should be contrasted with those for regular self-similar fractals~\cite{rammal1983nature,domany1983solutions,vanVeen2016, salvati2026emergence}: even if embedded into 2D space, a subextensive number of delocalized modes exist.
In our agglomerates, the geometrical disorder localizes all the eigenstates.

\begin{figure}
    \centering
    \includegraphics[width=0.47\linewidth]{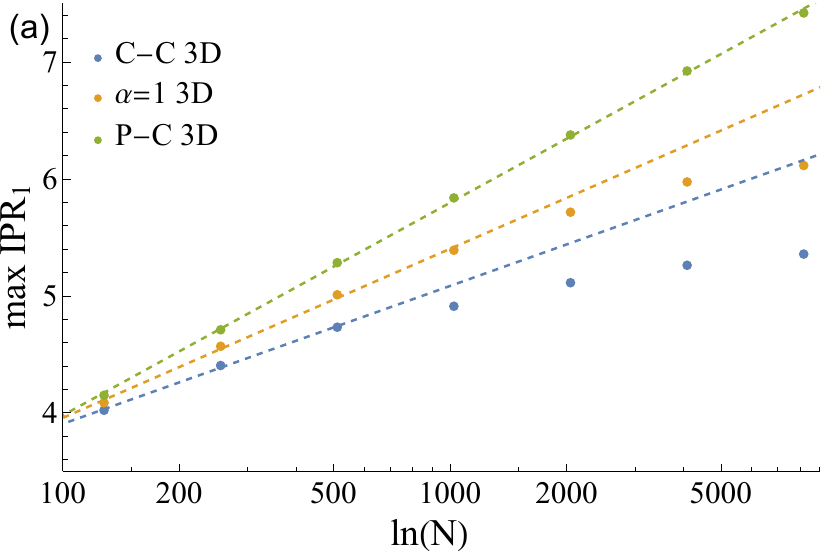}
    \includegraphics[width=0.47\linewidth]
    {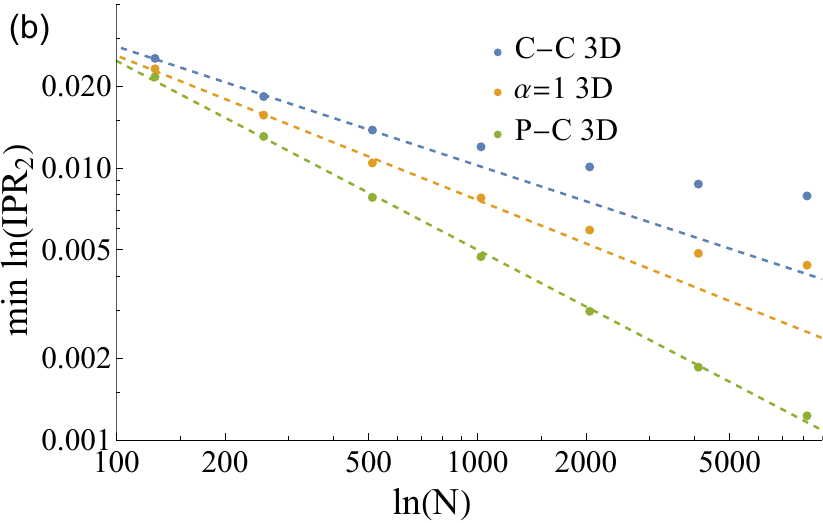}
    \includegraphics[width=0.49\linewidth]{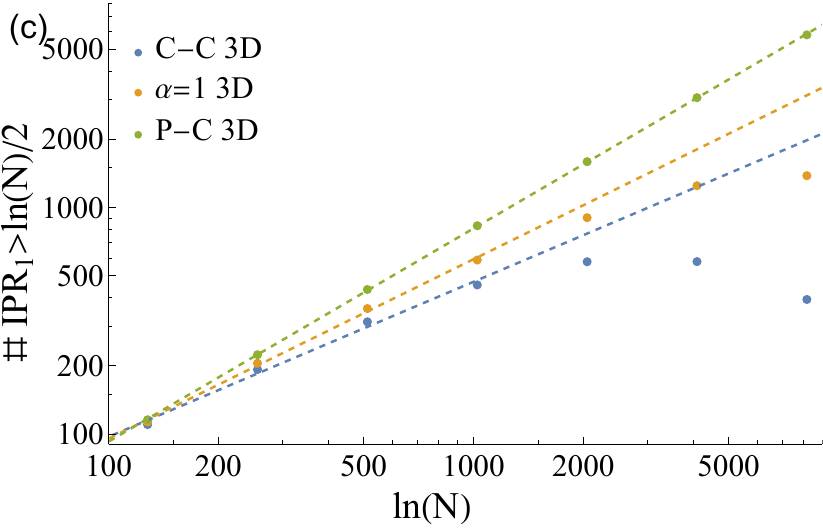}
    \includegraphics[width=0.49\linewidth]
    {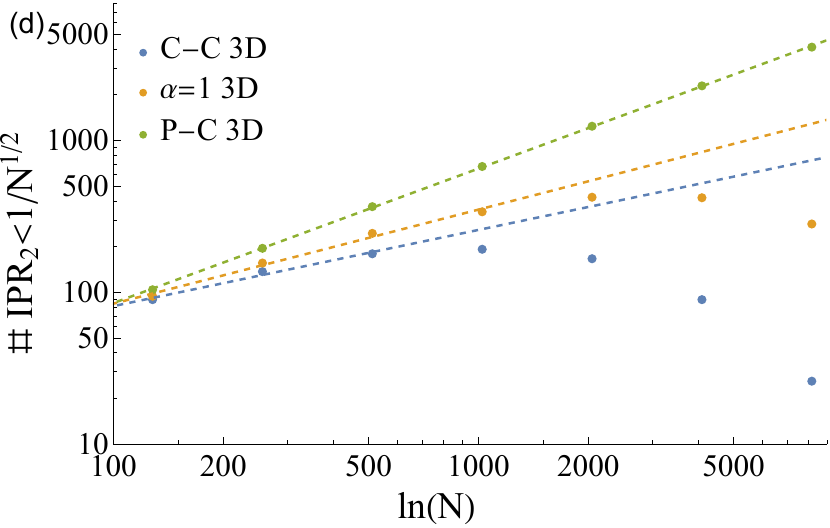}
    \caption{
        IPRs from exact diagonalisation for 3D agglomerates: C-C, \(\alpha=1\), P-C.
        Dashed lines are guides for the eyes.
        (a) Maximum \(\IPR_1\) averaged over \(64\) realizations per size.
        Eigenstates of P-C agglomerates are critical with \(\tau \approx 0.79\).
        (b) Minimum \(\IPR_2\) averaged over \(64\) realizations per size.
        Eigenstates of P-C agglomerates are critical with \(\tau \approx 0.69\).
        (c) Number of states with \(\IPR_1>(\ln{N})/2\) 
        (d) Number of states with \(\IPR_2 < \sqrt{N}\)
        }
    \label{fig:ipr3D}
\end{figure}

\sect{Localization--non-enrgodic transition in 3D}
In three dimensions, the spectral properties are qualitatively different.
In Figs.~\ref{fig:ipr3D}(a) and~(b), we show exact diagonalization results obtained in the same way as for the 2D agglomerates for the same system sizes.
Now, while for C-C and \(\alpha=1\) agglomerates our results resemble the 2D case, for the P-C ones, there is no evidence of saturation of the IPRs corresponding to the least localized states.
To be precise, our numerics suggest multifractal behavior with exponents \(\tau \approx 0.79\) (\(\IPR_1\)) and \(\tau \approx 0.69\) (\(\IPR_2\)).
This suggests the presence of non-localized, critical eigenmodes in the spectrum.

To estimate their number, we analyzed the scaling of the fraction of eigenstates above \(\IPR_1 > (\ln{N})/2\) and below \(\IPR_2 < \sqrt{N}\) cutoffs, see  Figs.~\ref{fig:ipr3D}(c) and~(d). For C-C and \(\alpha=1\) cases,  such modes quickly disappear with increasing \(N\), whereas for P-C structures their number scales as \(\approx N^{0.94}\) and \(\approx N^{0.89}\).
Therefore, even in the P-C case, the fraction of the states which we suspect to be critical tends to zero with \(N \to \infty\).

As we discussed above, for P-C agglomerates, the spectrum has localized and non-ergodic states intermixed, and the fraction of the latter vanishes with \(N\), making standard sparse diagonalisation ineffective in detecting the non-ergodic states.
To address this issue, we employed several approaches to detect non-localized states: wave-packet spreading, Green's functions, and sparse diagonalization with postselection.
In the first case, we consider unitary evolution of an initial state \(\psi_0\) localized on a single random site
\begin{gather}
    \label{eq:ue}
    \psi(t) = e^{-i \mh t} \psi_0.
\end{gather}
and calculate time-dependent IPRs from \(\psi(t)\).
One advantage of this approach is that the evolution does not require diagonalization and is computationally feasible for large system sizes. 
The other advantage is that our initial state typically overlaps with both localized and non-localized/non-ergodic modes (if present), and evolution up to sufficiently long times can reveal the non-localized states~\cite{dominguez2019aubry,vakulchyk2019wave}.
Notably, \(\IPR_2\) is dominated by localized modes and is not an effective probe for our purposes.
However, \(\IPR_{0,1}\) are sensitive to the existence of delocalized non-ergodic states in the spectrum and are instrumental in revealing the non-localized states (see End Matter).
For convenience of presentation, instead of \(\IPR_0\) we used \(N-\IPR_0/\ln{(1/\epsilon)}\) which can be interpreted as a number of ``visited sites''.

\begin{figure*}[t]
    \centering
    \includegraphics[width=0.3\linewidth]{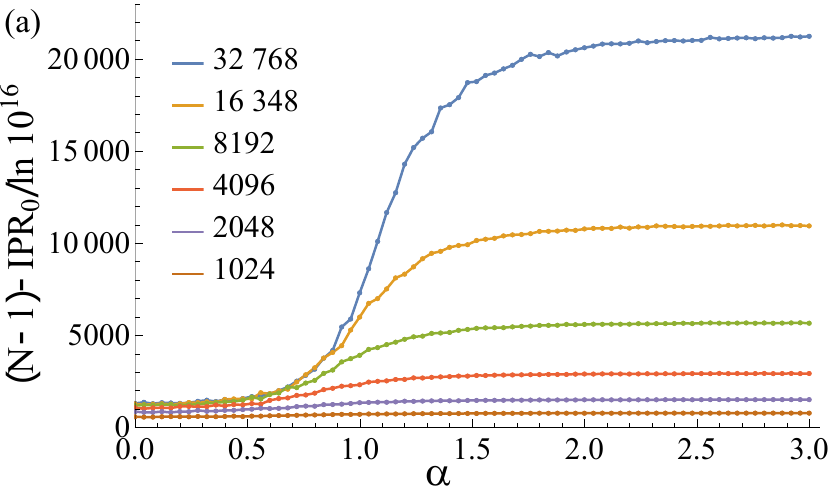}
    \includegraphics[width=0.3\linewidth]
    {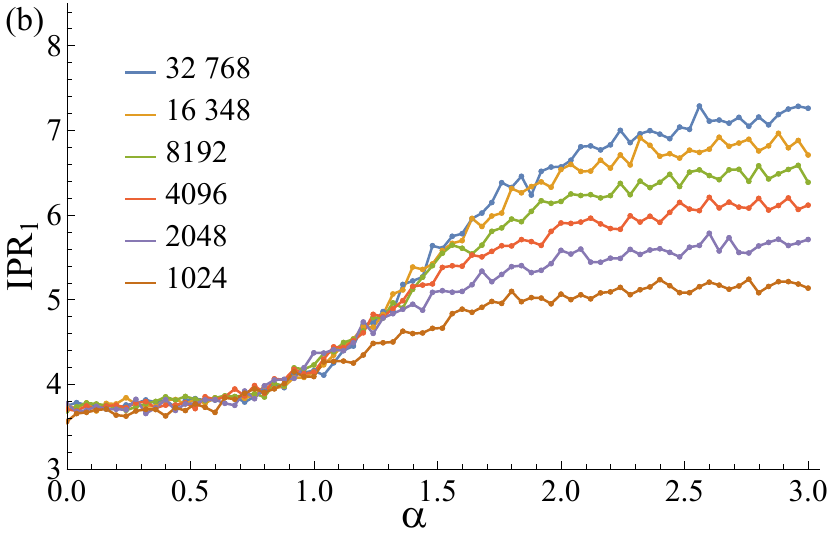}
    \includegraphics[width=0.3\linewidth]
    {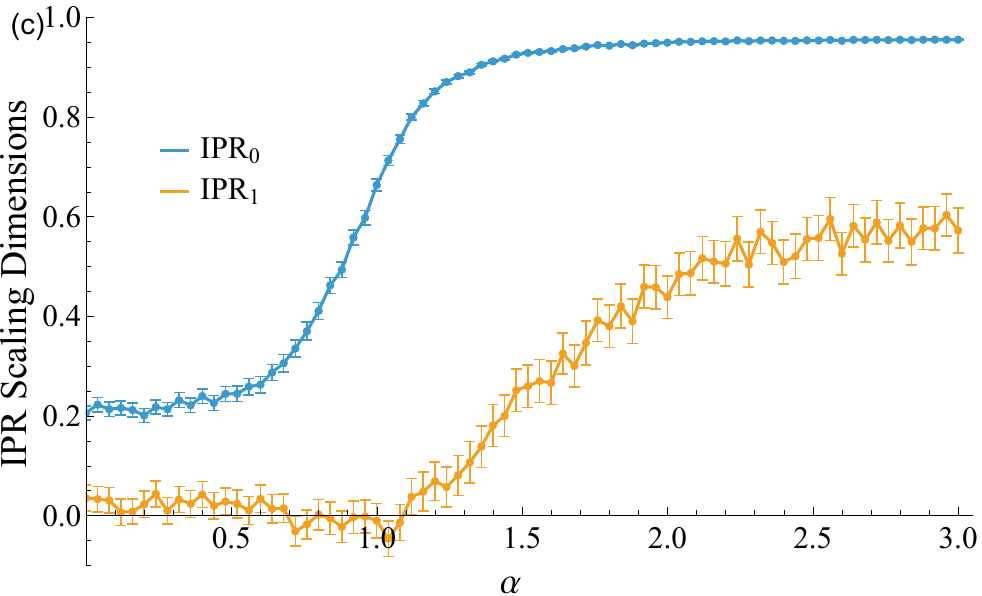}
    \caption{
        Time-evolved wavefunction from a random initial site of a 3D agglomerate.
        For each \(\alpha\) and size, we used \(20\) agglomerates and \(14\) initial states.
        (a) Number of ``visited sites'' related to \(\IPR_0\) (the vertical axis indicates the number of sites where the wave function amplitude is $\gg \epsilon^{1/2}=10^{-8}$) and 
        (b) \(\IPR_1\) at final times \(T=250 N/1024\).
        Both start scaling with system size for $\alpha \gtrsim 1.5$, manifesting the localized--non-ergodic transition in \(\alpha\).
        (c) Scaling dimensions of \(\IPR_{0,1}\) with statistical error bars.
        Finite values of scaling dimension of \(\IPR_0\) for $\alpha \lesssim 1.5$ are an artifact of the fitting procedure, see Supplementary Materials.
}
    \label{fig:time}
\end{figure*}

We computed the \(\IPR_{0,1}\) from the wavefunction~\eqref{eq:ue} for agglomerates of sizes  \(2^L, L=10\dots 15\) and \(\alpha \in [0,3]\) with \(76\) equidistant points.
This range of \(\alpha\) interpolates between sparse C-C agglomerates, where all eigenstates are localized according to Fig.~\ref{fig:ipr3D}, and P-C agglomerates, where some non-ergodic states are expected.
We set the final time of the evolution \(T \propto N\), in units of the inverse hopping (see End Matter for details).
The results were averaged over \(20\) random initial sites for each of the \(14\) agglomerate realizations.
We observed apparent evidence for a transition between the low-\(\alpha\) localized phase and high-\(\alpha\gtrsim 1.5\) phase, where non-ergodic fractal modes exist; see Fig.~\ref{fig:time}.
For \(\alpha \gtrsim 1.5\), \(\IPR_0\) is clearly scaling with \(N\); \(\IPR_1\), while noisier, also scales.

Another probe that addresses frequency rather than the time domain is the Green's function (details are presented in the End Matter).
We analyze the solution of the linear equation
\begin{gather}
    \label{eq:green}
    (E - \mh)_{ij} G_{j0}(E) = \delta_{i0}
\end{gather}
for \(E=0.57\) in our study (other energies, away from the spectrum boundary and DOS singularities, show similar results) and randomly chosen initial site \(0\).
The results are remarkably similar to those of the time-evolution method, see Fig.~\ref{fig:green}. 

Finally, we used sparse diagonalization with postselection to detect non-ergodic states, see Supplementary Material.
We obtained results similar to the ones based on the time-evolution and the Green's function methods, see Supplementary Fig.~S2.

We see that the behavior proposed above is confirmed using various numerical techniques.
For $\alpha \lesssim 1.5$, all the states of the tight-binding model on 3D agglomerates are localized.
However, for $\alpha \gtrsim 1.5$, there is a subextensive number of critical modes in the spectrum.
Thus, the transition between localized and non-ergodic regimes can be reported.
The numerical results above indicate that the transition occurs in the range \(1.3<\alpha_c<1.5\) near the inflection point of $\Df(\alpha)$ dependence [see Fig.~\ref{fig:fractals}(f)] where the fractal dimension smoothly saturates at the P-C limit.



\begin{figure}
    \centering
    \includegraphics[width=0.8\linewidth]{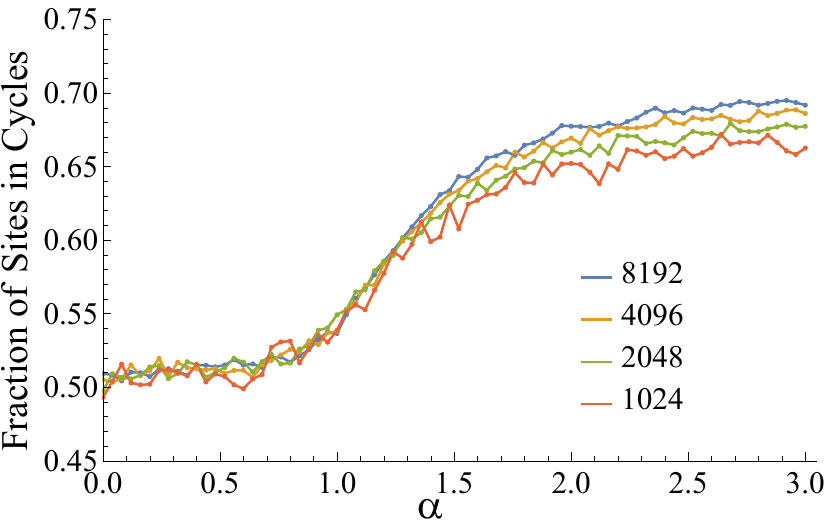}
    \caption{
        Fraction of sites on cycles as a function of \(\alpha\) for 3D agglomerates of size \(N=2^{10 \text{---} 13}\).
        Note the increase of the sites on cycles around \(\alpha \gtrsim 1.5\).
    }
    \label{fig:cycles}
\end{figure}

\begin{figure}[b]
    \centering
    \includegraphics[width=0.7\linewidth]{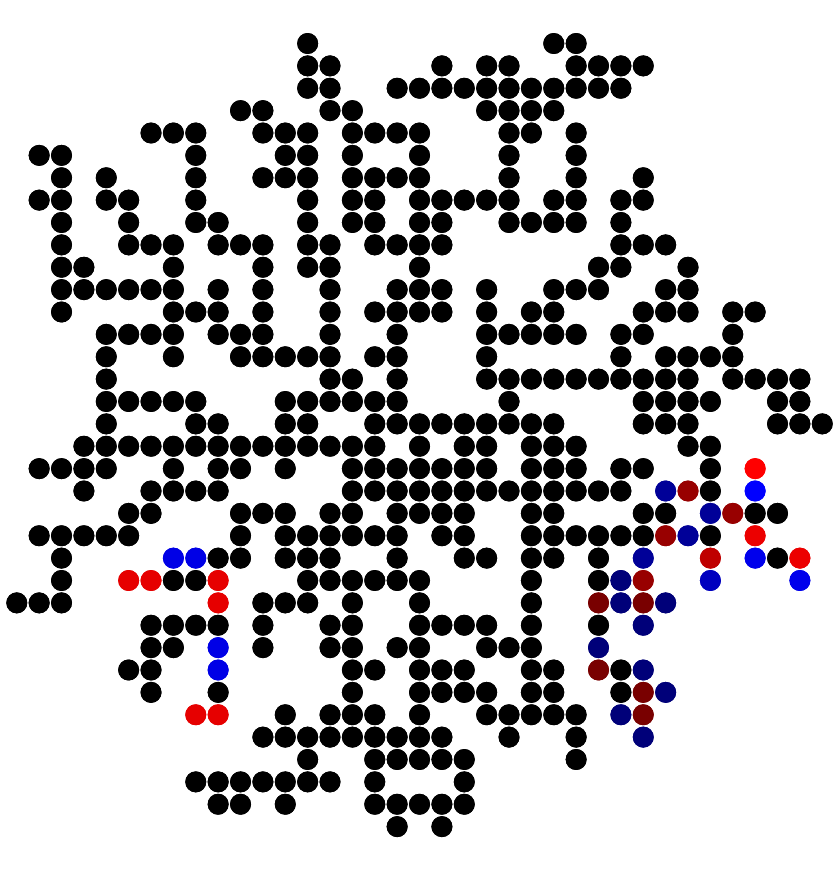}
    \caption{
        2D P-C agglomerate with \(N=512\). Spatially separated compact localized states with eigenvalue $E=1$ (left) and \(E=-1\) (right).
    }
    \label{fig:wfe1}
\end{figure}

\sect{Geometric aspect of localization--non-ergodic transition}An important characteristic of agglomerates related to their fractal dimension and geometrical structure is the fraction of sites being part of cycles or residing on ``dendrites'' (see Fig.~ \ref{fig:fractals}).
Indeed, we can split the sites of an agglomerate into two groups: those belonging to 1D ``dendrites'' and those belonging to cycles of the adjacency graph of the agglomerate, sites on cycles, for brevity.
In Fig.~\ref{fig:cycles} we report the fraction of sites on cycles for clusters with \(N =2^{10 \text{---} 13}\) sites and \(\alpha \in [0,3]\).
We see that for smaller \(\alpha\) an agglomerate mainly retains its ``hairy'' geometry: roughly half the sites belong to the dendrites.
For larger \(\alpha\gtrsim 1.5\), the number of sites on the cycles dominates and slowly grows with \(N\), corresponding to a denser ``globular'' structure, with fewer and shorter dendrites.
Importantly, there is a notable qualitative similarity between the results shown in Fig.~\ref{fig:cycles} and the IPRs scaling \(\alpha\)-dependence, see Figs.~\ref{fig:time} and \ref{fig:green} and Supplementary Fig.~S2.
In agglomerates with an abundance of dendrites, that are quasi-1D in nature, (smaller \(\alpha\)) localization is favored, while for larger \(\alpha\) dense, 3D-like structures dominate, responsible for a fraction of critical eigenstates in the spectrum.

\sect{Summary and outlook}
We proposed an algorithm generating fractal agglomerates with tunable dimensions and used it to investigate the properties of the conventional tight-binding model on the agglomerates.
We observed that for agglomerates embedded into 2D space, all the eigenmodes are localized, as expected from naive arguments.
However, in three spatial dimensions, we identified a transition in fractal dimension between a localized regime and a non-ergodic one, with a sub-extensive number of critical modes in the spectrum.
We further connected the transition with the change of the geometric structure of the agglomerates.
Moreover, in both 2D and 3D cases, the complex geometry of the agglomerates is responsible for singularities in the density of states, which are caused by an extensive number of compact localized states mainly residing on the structures' dendrites.

Our study connects physics of regular fractals, where exact self-similarity leads to a sub-extensive number of extended modes, and disordered integer-dimension structures, where the eigenmodes are typically localized or extended.
Our findings can also be important for understanding the nature and properties of agglomerates in multiple real-life physical systems.

\sect{Acknowledgements}
We are grateful to Boris Altshuler, Keith Slevin, Tilen \v{C}ade\v{z} and Sergej Flach for valuable discussions. 
AA and OU acknowledge financial support from the Institute for Basic Science (IBS) in the Republic of Korea through the Project No. IBS-R024-D1.
SK acknowledges financial support from the Institute for Basic Science (IBS) in the Republic of Korea through the YSF Project No. IBS-R024-Y3.

\bibliography{flatband,general,mbl,ref}

\clearpage

\begin{center}

\LARGE \bf End Matter
    
\end{center}

\sect{Degenerate states with \(E=\pm1\)}
Typical \(E= \pm 1\) eigenstates are shown in Fig.~\ref{fig:wfe1}.
Note that they come in pairs, i.e., if there is a state with $E=1$ made of the same-sign dimers, there is a complementary state with $E=-1$ made of alternating-sign dimers. 
Similarly to the \(E=0\) eigenstates shown in Figs.~\ref{fig:degen}(d)-(f) they also reside on dendrites of the agglomerates,

\begin{figure}
    \centering
    \includegraphics[width=0.7\linewidth]{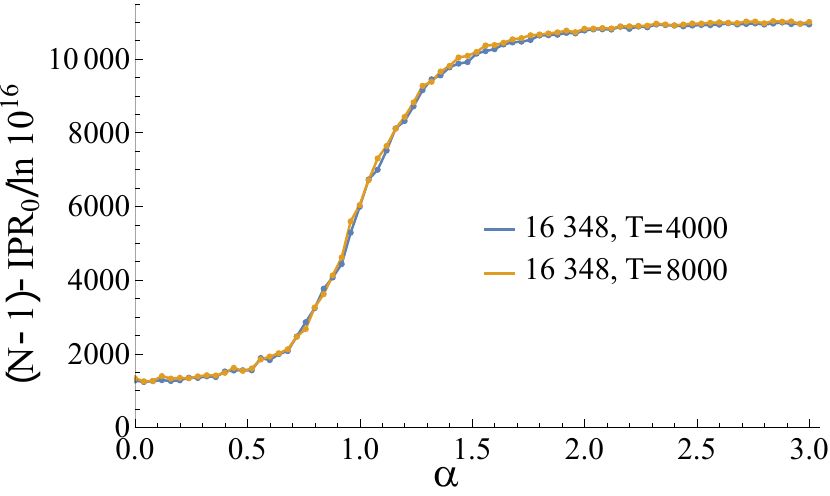}
    \caption{
        Number of visited sites, defined in the main text, for the P-C agglomerate of size \(N=2^{14}\) and final times \(T=4000\) (blue) and \(T=8000\) (orange) (in units of hopping).
        Doubling the final time \(T\) does not lead to appreciable differences in the number of visited sites.
        }
    \label{fig:twot}
\end{figure}

\begin{figure*}
    \centering
    \includegraphics[width=0.3\linewidth]{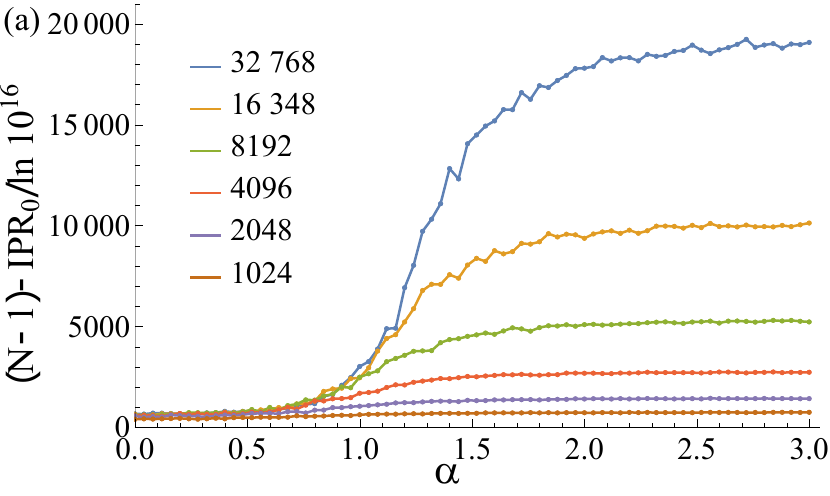}
    \includegraphics[width=0.3\linewidth]{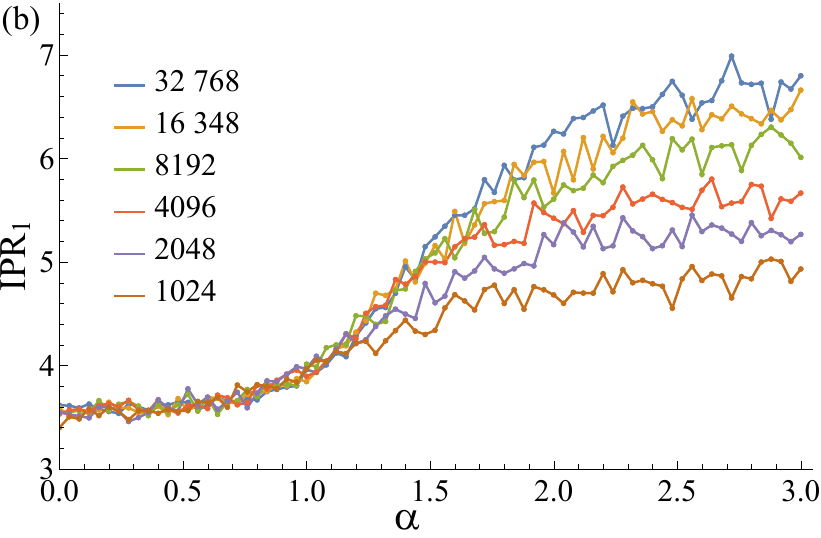}
    \includegraphics[width=0.3\linewidth]{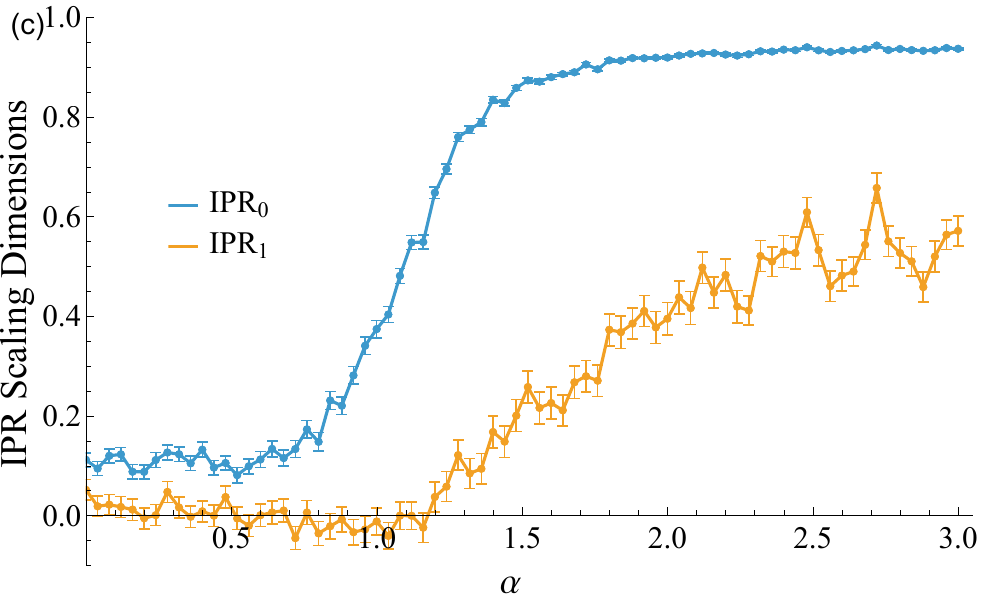}
    \caption{
        Localization properties based on IPRs computed from the normalized real-space Green's function \(G_{j0}(E)\) at energy \(E=0.57\).
        Clear signatures of non-localized states are present for \(\alpha \gtrsim 1.5\).
        (a) number of sites where Green's function has a non-negligible value $\gg \epsilon^{1/2}=10^{-8}$,
        (b) \(\IPR_1\),
        (c) scaling exponents of \(\IPR_0\) (blue) and \(\IPR_1\) (orange) vs \(\alpha\).
        For each \(\alpha\) and agglomerate size, we  averages over \(512\) values: \(16\) realizations of agglomerates times \(32\) initial sites 0.
        }
    \label{fig:green}
\end{figure*}

\sect{Time evolution}
For the unitary evolution on the 3D agglomerates, we used long enough evolution times, such that further increase of the final time does not change the results significantly.
Namely, single-site initial states were evolved unitarily until the time \(T\) at which \(\IPR_0\) and \(\IPR_1\) almost saturate.
In more detail, it means that the difference in IPRs on long time scales becomes of the order of the typical short-time fluctuation. 
Additionally, \(T\) was scaled with the size of the agglomerate \(N\).
E.g., for $N=2^{10}$ we used $T=250$ and $T=500$, for $N=2^{11}$ -- $T=500$ and $T=1000$, and for \(N=2^{14}\) --- \(T=4000\) and \(T=8000\). 
We ensured that the difference between the IPRs at these 2 times was of the order of the typical fluctuation.
As an illustration, we show in Fig.~\ref{fig:twot} that $T=4000$ and $T=8000$ lead to almost indistinguishable curves $\IPR_0$ for $N=2^{14}$ agglomerates.

\sect{Choice of observables}
The choice of \(\IPR_{0,1}\) as observables can be justified based on the following intuitive observation:
Let us consider \(\psi = 1/\sqrt{2} (\psi_\mathrm{loc} + \psi_\mathrm{deloc})\), where \(\psi_\mathrm{loc}\) is nonzero only on one site and \(\psi_\mathrm{deloc}\) is fully delocalized.
Then, it is straightforward to estimate \(N - \IPR_0/\ln{\epsilon} \approx N\) (the state is present everywhere), \(\IPR_1 \approx (\ln{2})/2 + \ln{N}\), and \(\IPR_2 \approx 1/4 + \mathrm{O}(1/N)\).
Both \(\IPR_{0,1}\) show respective $N$ and \(\ln{N}\) scaling with systems size, while \(\IPR_2\) saturates to a constant.
From this trivial example, one can see that \(\IPR_2\) is dominated by the overlap of the initial state with localized modes, while both \(\IPR_0\) and \(\IPR_1\) are sensitive to the presence of non-localized modes.

\sect{Green's function calculations} 
The normalized Green's function \(G\)~\eqref{eq:green} can be used to compute IPRs~\cite{von1996interaction,thongjaomayum2020multifractal}.
Typically, Green function~\eqref{eq:green} is controlled by eigenstates closest to energy \(E\) as follows from the spectral decomposition.
However, if those eigenstates are (compactly) localized and the site \(0\) is not in their support or sufficiently far away, other, non-localized eigenstates whose support includes \(0\) might contribute significantly to \(G\).
Therefore, using \(G\) to compute IPRs might address the issue of the intermixed spectrum of the agglomerates.
We observe that notably, \(\IPR_1\) shows a stronger dependence on the choice of initial site and overlap with localized states than \(\IPR_0\).
Therefore \(\IPR_0\) is a more reliable probe of delocalized/non-ergodic eigenstates with energies close to \(E\).
Still, we observed qualitatively similar results for both IPRs as shown in Fig~\ref{fig:green}.
Furthermore, these results both align well with our findings based on the unitary evolution shown in Fig.~\ref{fig:time}.
Both probes support the emergence of critical eigenmodes for $\alpha \gtrsim 1.5$ responsible for the subdiffusive wave packet spreading across the agglomerate.

\clearpage

\onecolumngrid

\newpage

\centerline{\large{\bf {Supplementary material to accompany the article}}} 

\vskip1mm

\centerline{\large{\bf {``Localisation--non-ergodic transition in controllable-dimension fractal networks from }}}

\centerline{\large{\bf {diffusion-limited aggregation''}}}

\vskip4mm

\centerline{Oleg I. Utesov, Alexei Andreanov, Tomasz Bednarek, Alexandra Siklitskaya, and Sergei V. Koniakhin} 

\vspace{1cm}

\counterwithin*{figure}{part}
\counterwithin*{equation}{part}
\stepcounter{part}
\renewcommand{\thefigure}{S\arabic{figure}}
\renewcommand{\thetable}{S\arabic{table}}
\renewcommand\theequation{S\arabic{equation}}
\renewcommand\thesection{S\arabic{section}}

\vskip10mm
\twocolumngrid

\section{Generation of the agglomerates}

The two cases of P-C (particle-cluster) and C-C (cluster-cluster) agglomerates have different fractal dimensions, with the dimension of the former being larger than that of the latter.
These agglomerates are generated by merging a single site with a cluster (P-C) and merging two equally-sized clusters (C-C).
This suggests that by varying the size of the merging clusters, one can also tune the fractal dimension of the resulting agglomerate.
This is known as the \emph{aggregation kernel}~\cite{van1988scaling,schmitt2000multiple} which is the reaction rate in the Smoluchowski equation. For our purposes, we used $K(M_1,M_2) =M^\alpha_1 M^\alpha_2$, where $\alpha$ is the crucial parameter allowing us to tune the fractal dimensions.

In our study, the following algorithm generated the agglomerates.
For each final size \(N\), the procedure starts from \(N\) monomers and results in a single agglomerate of size \(N\).
Assume that at a given step we have agglomerates of the sizes \(N_1, N_2,\dots\)
First, based on the size of the agglomerates, we calculated the probabilities \(P_i \propto N_i^{\alpha}\).
After this kinetic Smoluchowski equation-inspired phase, we ran the geometric real-space simulation of the discrete-steps-on-the-grid process of the two agglomerates approaching and colliding.
The agglomerates are formed on a 2D or 3D square grid, and their spatial structures are stored as coordinates of all particles belonging to the agglomerate.
Before joining, we rotate both agglomerates randomly (with discrete angles of $\pi/2$ step). Their mass centers are placed at a distance larger than their radii.
The direction from the bigger agglomerate to the smaller one is also chosen randomly.
With respect to this direction, we calculate probabilities to move the smaller agglomerate along each of the Cartesian axes (2 or 3) at unit distance, so that on average, the agglomerate centers of mass approach one another.
The procedure of joining is successful if the agglomerates join or fail after the distance between the agglomerates starts increasing, i.e., if no collision takes place.
In the latter case, we choose new relative positions between the agglomerates and repeat until successful joining.

For \(\alpha \gg 1\), the largest agglomerate is chosen as the rule.
The remaining agglomerates remain monomers. Thus, the P-C agglomeration is realized.
For \(\alpha \ll -1\), the algorithm always tends to choose the smallest agglomerates.
So, after the monomer stage, all monomers are joined to form dimers.
After that, in the same logic, dimers are joined to the agglomerates of size \(4\), and so on.
This procedure is, in fact, a C-C procedure.

This algorithm resulted in agglomerates with tunable fractal dimension.
In terms of sparsity, the resulting structures covered the full range, from the cluster-cluster agglomerates (sparse limit) to the particle-cluster agglomerates (dense limit).
However, such generation of agglomerates is time-consuming and is a bottleneck in the present study.
The construction of the entire agglomerate dataset takes approximately \(50\) hours.
Generation time is \(\alpha\)-dependent, being the longest for \(\alpha \approx 1\).
For instance, \(N=2048\) agglomerate creation takes up to \(10\) seconds on a 2025 laptop-class workstation [performance scales with agglomerate size as \(O(N^2)\)].
For C-C and P-C limits, the duration was of the order of \(2\) seconds on the same computer.
The KD-tree optimization was employed for the attachment of a single particle to a large agglomerate.
The hardest case was specifically the attachment of a few-particle agglomerate to a many-particle one.
Therefore, the intermediate \(\alpha\)s are slowest: the worst case is when dimers attach to a single big agglomerate.

\section{Density of states in 2D}

\begin{figure}[t]
    \centering
    \includegraphics[width=0.8\linewidth]{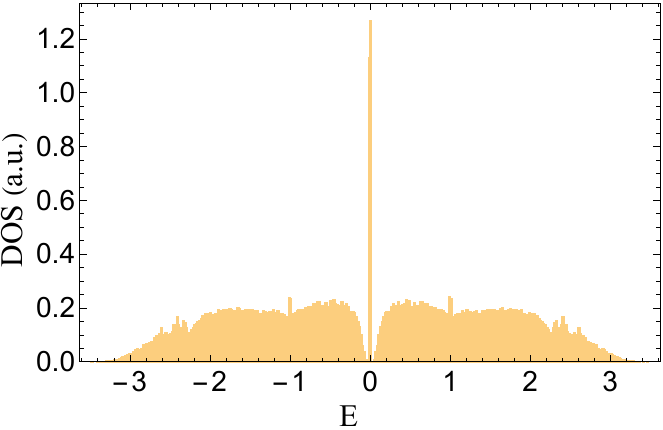}
    \caption{
        Density of states for 2D agglomerates with \(N=2^{12}\) and \(\alpha=0\), cf. Fig.~3(a) of the Main Text.
    }
    \label{fig:DOS2D}
\end{figure}

Here we present the density of states for the 2D agglomerates, complementary to the 3D one shown in Fig.~3(a) of the Main Text.
The DOS for \(\alpha=0\), averaged over \(64\) samples, is presented in Fig.~\ref{fig:DOS2D}.
Just like in the 3D case, there are peaks at $E=0, \, \pm1$.
Other $\alpha$ values result in qualitatively similar DOS.

\section{Sparse diagonalization with postselection}
\begin{figure*}
    \centering
    \includegraphics[width=0.3\linewidth]{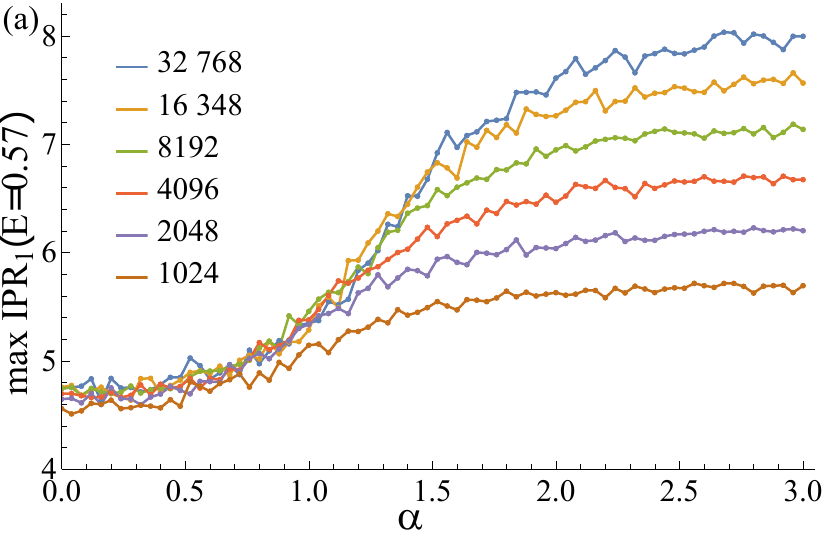}
    \includegraphics[width=0.3\linewidth]
    {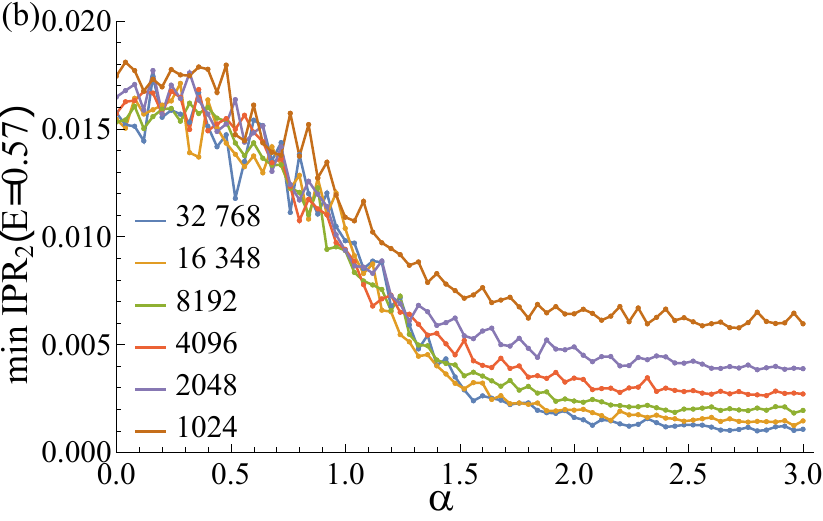}
    \includegraphics[width=0.3\linewidth]
    {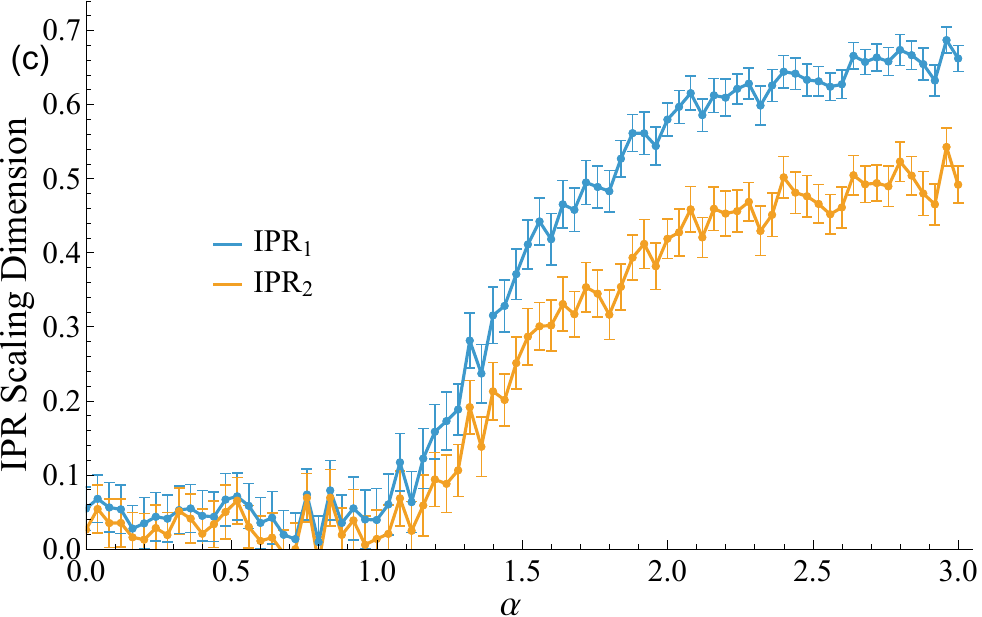}
    \caption{
        Average \(\IPR_{1,2}\) at \(E=0.57\) for 3D agglomerates computed using Arnoldi algorithm with postselection of least localized eigenstates out of \(21\) eigenstates.
        \(32\) realizations of agglomerates were used for each \(N\) and \(\alpha\).
        Clearly scaling of maximum \(\IPR_1\) and minimum \(\IPR_2\) with system size \(N\) is visible for \(\alpha \gtrsim 1.3\).}
    \label{fig:arnoldi}
\end{figure*}

Since both non-ergodic and localized states are intermixed in the spectrum, and the latter dominate, a typical eigenstate computed with sparse diagonalization is likely to be localized.  
To mitigate this, for agglomerates with \(N \leq 2^{15}\), we computed \(21\) states around target energies using the Arnoldi algorithm for each realization and chose the least localized one, with the largest \(\IPR_1\) and the smallest \(\IPR_2\).
The IPRs were averaged over \(32\) realizations for each \(N\) and \(\alpha\).
The results are shown in Fig.~\ref{fig:arnoldi}; they also suggest a transition around \(\alpha \approx 1.5\) between the phase where the entire spectrum is localized and the one where a vanishing fraction of modes is non-ergodic.
This result agrees with the other methods used in the Main Text.


\section{IPR scaling and \(\alpha\)}

Here we discuss particular fits that lie behind the IPRs' scaling dimensions shown in the Main Text and in the Supplemental Material, obtained using the time evolution [Fig.~6(c)], the Green's function [Fig.~9(c)], and Arnoldi [Fig.~\ref{fig:arnoldi}(c)] methods. 

\begin{figure*}
    \centering
    \includegraphics[width=0.4\linewidth]{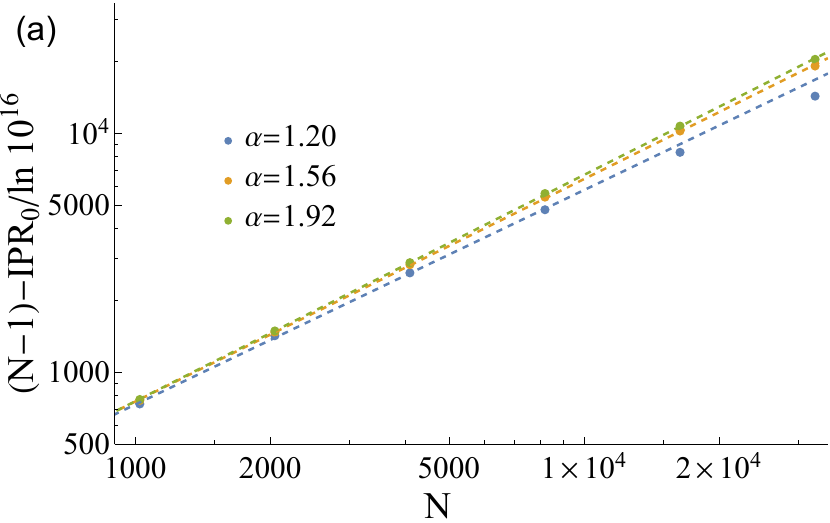}
    \hspace{1cm}
    \includegraphics[width=0.4\linewidth]{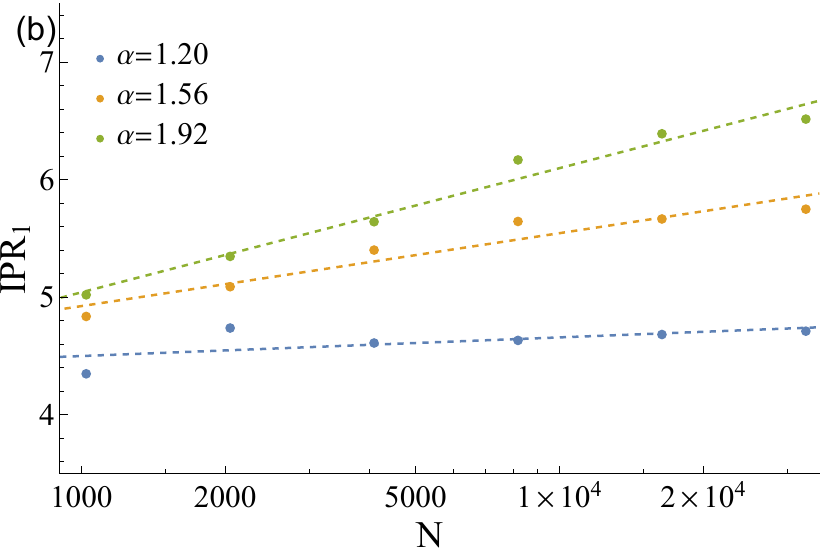}
    \centering
    \includegraphics[width=0.4\linewidth]{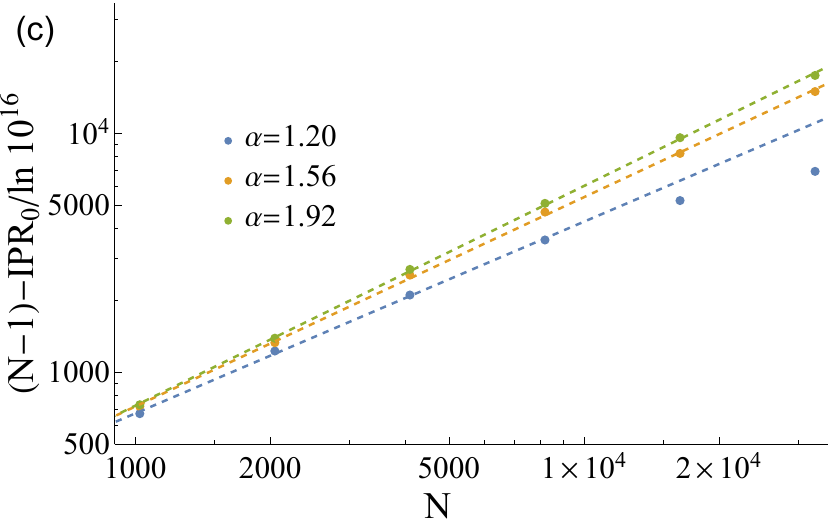}
    \hspace{1cm}
    \includegraphics[width=0.4\linewidth]{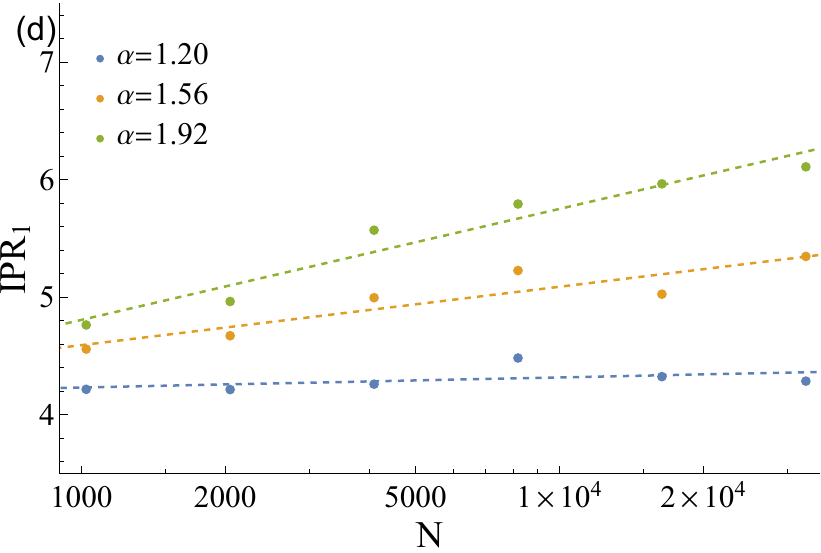}
    \centering
    \includegraphics[width=0.4\linewidth]{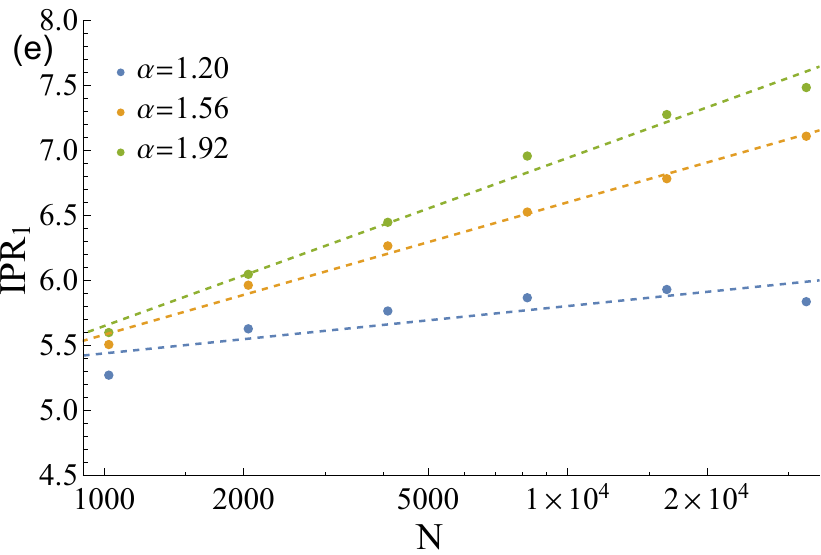}
    \hspace{1cm}
    \includegraphics[width=0.4\linewidth]{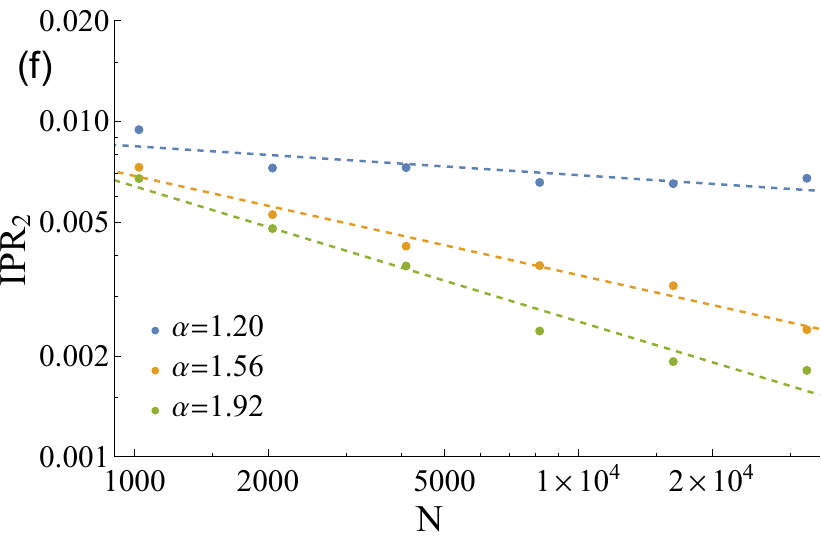}
    \caption{
        Scaling of IPRs computed with the 3 methods: unitary evolution, Green's function, and sparse diagonalization with postselection, for three select representative values of \(\alpha\).
        (a) and (b) (cf. Fig. 6 of the Main text), unitary evolution: 
        \(\alpha=1.2\) -- apparent localization; \(\alpha=1.56\) and \(1.92\) -- non-ergodic, multifractal eigenstates with \(\tau_0 \approx 0.93\), \(\tau_1 \approx 0.27\), and \(\tau_0 \approx 0.95\), \(\tau_1 \approx 0.46\) respectively.
        (c) and (d) (cf. Fig. 9 of the Main text), Green's function method: the results are remarkably similar. \(\alpha=0.76\) -- localization; \(\alpha=1.56\) and \(1.92\) -- non-ergodic behavior with \(\tau_0 \approx 0.88\), \(\tau_1 \approx 0.22\) and \(\tau_0 \approx 0.92\), \(\tau_1 \approx 0.41\) respectively.
        (e) and (f) (cf. Fig.~\ref{fig:arnoldi}), sparse diagonalization with postselection at \(E=0.57\) agrees with the two other methods.
        }
    \label{fig:scaling}
\end{figure*}

Notably, once again, we observe a good agreement among all three methods, see Fig.~\ref{fig:scaling}.
For \(\alpha \gtrsim 1.5\), we report the non-ergodic phase with a fraction of multifractal eigenstates.
For smaller \(\alpha\), there is a localized phase, where IPRs calculated by all the methods saturate with system size.
For sparse diagonalization with postselection, we demonstrate absence of scaling with system size, i.e. localization, for \(\alpha =1.2\) and multifractal eigenfunctions for \(\alpha = 1.56\): \(\tau_1 \approx 0.44\) and \(\tau_2 \approx 0.30\), and for \(\alpha = 1.92\): \(\tau_1 \approx 0.56\) and \(\tau_2 \approx 0.4\).

\end{document}